\documentclass[prb,preprint,url]{revtex4}

\usepackage[toc,page]{appendix}
\usepackage{hyperref}
\usepackage{graphicx,pdflscape}
\usepackage{color,subfigure}
\usepackage{bm}
\usepackage{alltt,dsfont}
\usepackage{appendix}
\usepackage{simpler-wick}
\usepackage{amsmath,amssymb}
\usepackage{placeins,float}
\usepackage{atveryend}
\makeatletter
\let\origcitation\citation
\AtEndDocument{\def\mycites{}%
  \def\citation#1{\g@addto@macro\mycites{#1^^J}\origcitation{#1}}}
\AtVeryEndDocument{\newwrite\citeout\immediate\openout\citeout=\jobname.cit
  \immediate\write\citeout{\mycites}\immediate\closeout\citeout}
\makeatother

\newcommand{\llabel}[1]{  \label{#1} }

\newcommand {\apgt} {\ {\raise-.5ex\hbox{$\buildrel>\over\sim$}}\ }
\newcommand {\aplt} {\ {\raise-.5ex\hbox{$\buildrel<\over\sim$}}\ }
\newcommand{\lessim}{\aplt}
\newcommand{\gssim}{\apgt}

\newcommand{\iden}{ \mathds{ 1}}
\newcommand{\G}{{\cal{G}}}
\newcommand{\GH}{{\bf g}}
\newcommand{\GHI}{\GH^{-1}}

\newcommand{\tr}{{\text {Tr} }}
\newcommand{\V}{{\mathcal V}}

\newcommand{\X}[2]{X_{{#1}}^{#2}}
\newcommand{\si}{\sigma}
\newcommand{\sib}{\bar{\sigma}}
\newcommand{\tJ}{$t$-$J$  }
\newcommand{\beq}{\begin{eqnarray}}
\newcommand{\eeq}{\end{eqnarray}}
\newcommand{\barray}{\begin{eqnarray}}
\newcommand{\earray}{\end{eqnarray}}
\newcommand{\nn}{\nonumber}
\newcommand{\A}{\hat{{\mathcal A}}}

\renewcommand{\AA}{{{\mathcal A}}}

\newcommand{\disp}[1]{Eq.~(\ref{#1})}

\newcommand{\refdisp}[1]{Ref.~(\onlinecite{#1})}
\newcommand{\figdisp}[1]{Fig.~(\ref{#1})}

\newcommand{\up}{\uparrow}
\newcommand{\dn}{\downarrow}

\newcommand{\chem}{{\bm \mu}}

\renewcommand{\cap}{\widetilde{\mu}}

\usepackage{color}
\usepackage{bm}
\usepackage{alltt,dsfont}
\usepackage{appendix}
\usepackage{amsmath,amssymb}
\usepackage{hyperref}
\hypersetup{
    colorlinks,
    citecolor=red,
    filecolor=cyan,
    linkcolor=blue,
    urlcolor=magenta
}

\usepackage{atveryend}
\makeatletter
\let\origcitation\citation
\AtEndDocument{\def\mycites{}%
  \def\citation#1{\g@addto@macro\mycites{#1^^J}\origcitation{#1}}}
\AtVeryEndDocument{\newwrite\citeout\immediate\openout\citeout=\jobname.cit
  \immediate\write\citeout{\mycites}\immediate\closeout\citeout}
\makeatother

\begin{document}
\title{ Extremely Correlated Fermi Liquid theory    of the   $t$-$J$ model in 2 dimensions: \\Low Energy properties}
\author{ B. Sriram Shastry  }
\author{Peizhi Mai  }
\affiliation{Physics Department, University of California,  Santa Cruz, Ca 95064 }
\date{August 24, 2017}
\begin{abstract}
Low energy  properties of the metallic state of the 2-dimensional \tJ model are presented at various densities and temperatures for  second neighbor hopping $t'$, with signs that are   negative or positive corresponding to  hole or electron doping. The  calculation employs a closed set of equations for the Greens functions obtained from  the extremely correlated Fermi liquid theory.  These  equations, when used  in $d=\infty$ reproduce most of the known low energies features  of the  $U=\infty$  Hubbard model. In  2-dimensions we are able to study the variations due to the  superexchange J.  The resulting Dyson self energy is found to be  momentum dependent as expected. The  density and temperature dependent quasiparticle weight, decay rate and the   peak spectral heights over the Brillouin zone are calculated.  We also calculate the    resistivity,  Hall conductivity and  cotangent  of the Hall angle in experimentally relevant units. These display  significant   thermal sensitivity for density $n \gssim 0.8 $, signifying   an effective   Fermi-liquid temperature  scale which  is  { two or three orders of magnitude} below the bare bandwidth.  Flipping  the sign of the  hopping   $t'$, i.e.  studying hole versus electron doping,  is found to     induce a change in curvature of the temperature dependent resistivity from convex to concave at  low temperatures.  Our  results 
 provide  a natural route for  understanding the observed  difference in the temperature dependent resistivity of strongly correlated electron-doped and hole-doped  matter.
\end{abstract}
\maketitle

\section{Introduction}

The  \tJ model in 2-dimensions (2-d) has been argued to be  of  fundamental importance for understanding
strongly correlated matter, including the 
  high Tc superconductors\cite{PWA,tJ-review1}. Due to the  difficulties inherent in a strong coupling problem, very  few techniques are available for extracting its low temperature physics. Towards this end     we have recently developed  the extremely correlated Fermi liquid  (ECFL) theory \cite{ECFL,Pathintegrals}.  It is an  analytical method  for treating very strong correlations of lattice Fermions,  employing Schwinger's technique of functional differential equations togather  with several important added ingredients. While  further details can be found  in \refdisp{ECFL,Pathintegrals}, a brief summary of the main idea behind the ECFL theory seems appropriate.   We consider  the  Hubbard model with a large  interaction $U\-\to \-\infty$, and hence the name of the theory. A well known  expansion in the inverse powers of U leads   to the  \tJ model (defined below \cite{tJ-review1}).   Taking the infinite U limit  forces one to   abandon the conventional Feynman diagram based perturbation theory in $U$, and to make a fresh start. The ECFL theory starts with the graded Lie-algebra of the Gutzwiller projected, i.e. infinite-U limit Fermi operators Eqs.~(\ref{commutator1},\ref{commutator2}). This  leads to an exact functional differential  equation for the Greens functions, known as the Schwinger equation  of motion  Eq.~(\ref{EOM-42} or \ref{Min-1}). 
  In this equation,   a   parameter $\lambda$ is introduced; $\lambda$ is
   bounded in the range $\in [0,1]$ and  represents the evolution   from the free Fermi limit. We then  use a systematic expansion in the parameter $\lambda$, for solving the Schwinger  equations  perturbatively  in  $\lambda$. In this scheme we start with the uncorrelated Fermi gas at $\lambda=0$ and end up at the fully correlated projected Fermion problem at $\lambda=1$.  The scheme thus represents a generalization of the usual  perturbation theory for canonical Fermionic models,  in order to handle a non-canonical Fermionic problem such as   the \tJ model. 
      The context of interacting Bosons provides a useful parallel. In the well known  problem of representing spin $S$ variables  in terms of  canonical Bosons,  one uses  the   expansion parameter $\frac{1}{2S}$ with a similar range  $ \frac{1}{2S} \in [0,1]$. We may think of $\lambda$ as being  analogous to  the parameter $\frac{1}{2S}$ as shown in \refdisp{Pathintegrals}. The introduction of the parameter  $\lambda$ and the $\lambda$-expansion scheme thus enabled  are among the main technical advances introduced in the ECFL theory.

   This approach has been recently benchmarked \cite{Sriram-Edward,Edward-Sriram,WXD} against the numerically exact  results from the single impurity Anderson model, and   the    $d=\infty$ Hubbard model from    Dynamical Mean Field Theory (DMFT) \cite{badmetal,HFL}. These tests provide quantitative support to our general  scheme described below, especially for low energy response.  Our scheme has no specific limitation to $d=\infty$, and is expected to be reasonably accurate in any dimension $d>1$, including  2 dimensions, a case of great experimental importance due to the High Tc cuprate materials. It   is applied here to probe  the metallic state of the \tJ model in 2-d.   We  present results for the electron self energy, the spectral functions, the resistivity  the Hall constant and the Hall angle  at various temperatures and  electron density $n=N/N_s$ (number of electrons per site).  We also use the notation of  hole density $\delta =(1-n)$ instead of $n$, following the convention used in several experimental studies of doped  Mott systems..

We explore  various values of the parameters of the \tJ model, including the  second neighbor hopping, which turns out to play a very important role in determining the effective Fermi Liquid (FL) temperature scale. We investigate  the  resistivity due to mutual collisions of electrons at low temperatures, and its dependence on the parameters of the model.  We pay special attention to the resistivity since this easily  measured -but notoriously  hard to calculate object, reveals the lowest energy scale physics of charge excitations in metallic systems, and  therefore is of central importance.

\section{Methods}
In this section we summarize the equations used in the present calculation,  together with  the arguments leading to them- further details may be found in earlier papers on this theory\cite{ECFL,Sriram-Edward,Edward-Sriram,WXD}. In Sec.~(\ref{preliminaries}) the model is defined and the exact Schwinger-Dyson equations of motion are written out. In Sec.~(\ref{lambda-expansion}) the $\lambda$ parameter is introduced and the exact factorization of the Greens function into an auxiliary Greens function and a caparison function are noted. In   Sec.~(\ref{second-chem}) we summarize the  shift identities of the \tJ model. The shift transformation is a simple and yet important invariance of the  \tJ model  leading to important constraints on possible approximations. Within the $\lambda$ expansion,  this invariance obligates the introduction of a second chemical potential $u_0$, which is then treated as a Lagrange multiplier to be fixed through sum-rules.
In Sec.~(\ref{secondorderequations}) we collect the  equations of the second order  theory. In Sec.~(\ref{cutoff}) we  summarize the rationale for a high energy cutoff of the equations given in Sec.~(\ref{secondorderequations}). 

\subsection{The \tJ model preliminaries \label{preliminaries} }

The  \tJ model is a two component Fermi system on a lattice, defined on the restricted subspace of  three  local states, obtained by excluding all doubly occupied configurations.  The allowed states at a single site are  $|a\rangle$ with $a=0,\up,\dn$, and the double occupancy state $|\up \dn\rangle$ is removed by the (Gutzwiller) projection operator $P_G= \Pi_i (1-n_{i \uparrow} n_{i \downarrow})$. We use the Hubbard operators $\X{i}{a,b} = |a\rangle \langle b|$, which are expressible in terms of the usual Fermions $C_{i \si}, C^\dagger_{i \si}$  and the Gutzwiller projector $P_G$ as: 
\barray
\X{i}{\sigma 0}= P_G C^\dagger_{i \sigma} P_G; \;
\X{i}{0 \sigma }= P_G C_{i \sigma} P_G; \;
\X{i}{\sigma \sigma'}=  P_G C^\dagger_{i \sigma} C_{i \sigma'} P_G.
\earray
These obey the anti-commutation relations
\beq
\{\X{i}{0 \sigma_i}, \X{j}{\sigma_j,0} \} = \delta_{i,j} \left( \delta_{\sigma_i,\sigma_j} -{\sigma_i \sigma_j} \X{i}{\bar{\sigma}_i,\bar{\sigma}_j}\right) \label{commutator1}
\eeq
and the commutators
\beq
~[  \X{i}{0 \si_i}, \X{j}{\si_j \si_k}]=\delta_{i j} \delta_{\si_i \si_j} \X{i}{0 \si_k}; \;
~[  \X{i}{ \si_i 0 }, \X{j}{\si_j \si_k}]=- \delta_{i j} \delta_{\si_i \si_k} \X{i}{ \si_j 0}  \label{commutator2}.
 \eeq
The  Hamiltonian of the general \tJ model  $H_{tJ}$ is  
\barray
H_{tJ}&=& H_t+H_J, \nn \\
H_{t}&=& - \sum_{i j } t_{ij} \X{i}{\si 0}\X{j}{0\si} - \chem \sum_i \X{i}{\si \si}; \;\;
H_J=  \frac{1}{2} \sum_{ij} J_{ij} \left( \vec{S}_i . \vec{S}_j - \frac{1}{4} \X{i}{\si \si} \X{j}{\si' \si'}   \right), \label{hamiltonian}
\earray
where we sum over repeated spin indices.
Here $\chem$ is the chemical potential and  the spin is given in terms of the Fermions and the Pauli matrices $\vec{\tau}$  as usual $\vec{S} = \frac{1}{2} \X{i}{\si 0} \vec{\tau}_{\si \si'} \X{i}{0 \si'} $.
We will restrict in the following to nearest neighbor exchange  $J$,  and first (t) and second neighbor (t') hopping on a square lattice.

For the purpose of computing the Green's functions we  add     Schwinger  sources to the Hamiltonian;  the  commuting (Bosonic) potential $ \V$  couples to the   charge as well as spin  density.  These sources  serve to generate compact Schwinger equations of motion (EOM),  and are set to zero at the end. The zero source equations are usually termed as the Schwinger-Dyson equations.  In that limit we recover spatial and temporal translation invariance of the Greens function.  Explicitly we write
\barray
\A_S & = & \sum_i \int_0^\beta   \A_S(i, \tau) d \tau; \;\;\A_S(i, \tau) =   \V_{i}^{\si' \si}(\tau) \X{i}{\si' \si}(\tau), \llabel{sources}
\earray
and all time dependences are as in $Q(\tau)=e^{\tau H_{tJ}}Q e^{-\tau H_{tJ}}$.  The generating functional of Green's functions of the \tJ model is
\beq
Z[\V]\equiv \tr_{tJ} \ e^{-\beta H_{tJ}}  T_\tau \left( e^{- \A_S}  \right). \llabel{part1}
\eeq
it reduces to the standard partition function on turning off the indicated source terms. The Green's functions for positive  times $0 \leq \tau_j \leq \beta$,
 are defined  as usual:
\beq
\G_{\si \si'}(i  \tau_i, f  \tau_f) = -   \langle T_\tau \left( e^{- \A_S}   \X{i}{0 \si}(\tau_i) \X{f}{\si' 0}(\tau_f)\right)\rangle  .  \label{gdef}
\eeq
where for an arbitrary ${\cal Q}$ we define
\beq
\langle {\cal Q} \rangle \equiv  \frac{1}{Z} \tr_{tJ} \ e^{-\beta H_{tJ}}  T_\tau  \left( e^{- \A_S}  {\cal Q}\,\right)
\eeq
  We note that $n_\si$,  the number  of particles per site, is determined from the number sum rule:
\beq
n_{\si}= \G_{\si \si}(i \tau^-, i \tau), \label{number-sumrule}
\eeq 
and $\chem$ the chemical potential is fixed by this constraint. By taking the time derivative of \disp{gdef} we see that the Green's function satisfies the EOM
\beq
&&\partial_{\tau_i} \G_{\si_i \si_f}(i , f)= - \delta(\tau_i-\tau_f) \delta_{if} (1- \gamma_{\si_i \si_f}(i \tau_i) ) -   \langle T_\tau \left( e^{- \A_S} [H_{tJ}+ \A_S(i,\tau_i), \X{i}{0 \si_i}(\tau_i)] \ \X{f}{\si_f 0}(\tau_f) \right) \rangle  \nn \\
\label{EOM1} 
\eeq
where the local Green's function is defined as
 \beq\gamma_{\si_a \si_b}(i \tau_i) = \si_a \si_b \G_{\sib_b \sib_a}(i \tau_i^-, i \tau_i) \;, 
  \label{gamma-def}
 \eeq
 with the notation
 \beq
 \sib_i= - \si_i.
 \eeq

Using the Hamiltonian \disp{hamiltonian} and canonical relations Eqs.~(\ref{commutator1}, \ref{commutator2}) we find
\barray
&&[ H_{tJ},\X{i}{0 \si_{i}}]  =   \sum_j t_{ij} \X{j}{0 \si_i}  
 + \chem \X{i}{0 \si_i}   - \sum_{j \si_j} t_{ij} (\si_i \si_j)  \X{i}{\sib_i \sib_j} \ \X{j}{0 \si_j} + \frac{1}{2} \sum_{j \neq i} J_{ij} \ (\si_i \si_j)   \X{j}{\sib_i \sib_j}   \X{i}{0 \si_j},  \nn \\
 \label{commutator} 
\earray
and 
\beq
[\A_S(i \tau_i),\X{i}{0 \si_{i}}]= -\V_i^{\si_i\si_j} \X{i}{0 \si_j}. \label{aa}
\eeq
Substituting into \disp{EOM1}  and using the  free Fermi gas  Green's function:
\beq
 &&\GHI_{0, \si_i,  \si_j}(i \tau_i,  j \tau_j)=  \left\{  \delta_{\si_i \si_j} \left[\delta_{ij} (\chem- \partial_{\tau_i}) + t_{ij} \right]  - \delta_{ij} \V_i^{\si_i \si_j}(\tau_i)\right\} \delta(\tau_i-\tau_j),  \label{gnon} 
 \eeq
we obtain 
\beq
&&\GHI_{0, \si_i,  \si_j}(i \tau_i,  j \tau_j) \G_{\si_j \si_f}(j \tau_j , f \tau_f)=  \delta(\tau_i-\tau_f) \delta_{if} (1- \gamma_{\si_i \si_f}(i \tau_i) ) \nn \\
&& -  \sum_{j \si_j} t_{ij} (\si_i \si_j) \   \langle T_\tau \left(\X{i}{\sib_i \sib_j}(\tau_i)  \X{j}{0 \si_j}(\tau_i)  \ \X{f}{\si_f 0}(\tau_f) \right)\rangle    + \frac{1}{2 } \sum_{k \si_j} J_{ik}  (\si_i \si_j) \langle T_\tau \left(   \X{k}{\sib_i \sib_j}(\tau_i)   \X{i}{0 \si_j}(\tau_i)  \X{f}{\si_f 0}(\tau_f) \right) \ .\nn \\ \label{EOM-421} 
\eeq
 We next ``reduce'' the higher order Green's function to a lower one using the identity (valid for any operator ${\cal Q}$):
 \beq
 \langle T_\tau \X{i}{\si \si'}(\tau) {\cal Q} \rangle = \langle T_\tau \X{i}{\si \si'}(\tau) \rangle \, \langle T_\tau {\cal Q} \rangle -  \frac{\delta}{\delta \V_i^{\si \si'}(\tau)}  \langle T_\tau {\cal Q} \rangle,
 \eeq
 and rearranging terms we  obtain the fundamental Schwinger EOM:
 \beq
  \left(  \GHI_{0, \si_i,  \si_j}(i \tau_i,  j \tau_j) -  \hat{X}_{ \si_i \si_j}(i \tau_i, j \tau_j)-  {Y_1}_{ \si_i \si_j}(i \tau_i, j \tau_j)\right) 
  \times \G_{\si_j \si_f}( j \tau_j, f \tau_f) = \delta_{if} \delta(\tau_i-\tau_f)  \left( \delta_{\si_i \si_f} - \gamma_{\si_i \si_f }(i \tau_i) \right),  \nn \\ \label{EOM-42}
 \eeq
 where
we defined the functional derivative operator at site $i$ and time $\tau_i$
\barray
 {D}_{\si_i \si_j}(i \tau_i)& = &  \si_i \si_j { \frac{\delta}{\delta \V_i^{\sib_i \sib_j}(\tau_i)}} , \label{def-der}
 \earray
the composite derivative operator
\beq
&&\hat{X}_{\si_i \si_j}(i \tau_i, j \tau_j)=\delta(\tau_i-\tau_j) \times \left( - t_{ij} D_{\si_i \si_j}(i \tau_i)+ \delta_{ij} \sum_k \frac{1}{2} J_{ik} D_{\si_i \si_j}(k \tau_i ) \right),\llabel{xopdef2}
\eeq and  corresponding $Y_{1}$  as 
\beq
&&{Y_1}_{\si_i \si_j}(i \tau_i, j \tau_j)=- \delta(\tau_i-\tau_j) \times \left( - t_{ij} \gamma_{\si_i \si_j}(i \tau_i)+ \delta_{ij} \sum_k \frac{1}{2} J_{ik} \gamma_{\si_i \si_j}(k \tau_i) \right).\llabel{ydef2}
\eeq
By considering the spin, space and time variables as generalized matrix indices, we can symbolically write \disp{EOM-42} as   
\beq
\left(  \GHI_{0} -  \hat{X}-   {Y_1}\right). ~\G = \delta \  ( \iden  -  {\gamma}  ). \label{Min-1}
\eeq  

\subsection{ The $\lambda$ expansion and the auxiliary Greens function  \label{lambda-expansion} }

The main task  is to compute  solutions of the Schwinger-Dyson equation, i.e. the functional differential equation \disp{EOM-42} or \disp{Min-1}. If  symmetry-breaking, such as magnetism or superconductivity   is ignored, then  a liquid state ensues,  where we would like the  solution to connect continuously with the Fermi gas.
  For this purpose we seek guidance from standard Feynman-Dyson perturbation theory for canonical models.  The repulsive Hubbard model is an ideal example, where the corresponding Schwinger-Dyson   equation can be schematically written as:
\beq
\left(  \GHI_{0} -  U \delta/{\delta{\V}} -  U  G \right). G = \delta \   \iden.  \label{Min-Hub}
\eeq
Comparing with \disp{Min-1}, we see that the left-hand sides are of the same form,  but the right-hand sides  differ, in \disp{Min-1}  the local  Greens function $\gamma$ multiplies the delta function. In turn this extra term originates from the second (non canonical) term in the anti-commutator in \disp{commutator1}, and is therefore  the signature term  of extremely strong correlations.

Within the Schwinger viewpoint of \disp{Min-Hub}, we can view the skeleton graph  perturbation theory (Feynman-Dyson) as an iterative scheme in $U$, i.e. using the $n^{th}$ order results to generate the $(n+1)^{th}$ order terms by functional differentiation.
In the ECFL theory 
 the iterative scheme used is defined by generalizing \disp{Min-1} to
\beq
\left(  \GHI_{0} -  \lambda \hat{X}-  \lambda {Y_1}\right). ~\G = \delta \  ( \iden  - \lambda {\gamma}  ). \label{Min-22}
\eeq
The explicit solutions in the ECFL theory start  from this basic equation. 
More explicitly, in \disp{Min-22} the exact    \disp{EOM-42} is  generalized to include the $\lambda$ parameter\footnote{In \refdisp{Pathintegrals} we have noted an important generalization of these commutators to include a continuous parameter $\lambda \in[0,1]$, thus defining the so called $\lambda$ Fermions.  Using them one can systematically obtain the $\lambda$ expansion encountered below from these relations directly. Here we stick to  a simpler description with $\lambda$ introduced by hand, in the equations of motion below. }   by scaling $\hat{X}_{\si_i \si_j}, Y_{i \si_i \si_j}, \gamma_{\si_i \si_j} \to \lambda \hat{X}_{\si_i \si_j}, \lambda Y_{i \si_i \si_j}, \lambda \gamma_{\si_i \si_j}$.
 The starting point for the iteration is $\lambda=0$,  corresponding to  the Fermi gas. As  we iterate towards  $\lambda=1$, \disp{Min-22} reduces to the exact equation \disp{Min-1}.  The Gutzwiller projection is fully effective only at the end point of the iterative scheme $\lambda=1$, while for intermediate values of $\lambda$, we have only a partial reduction of the number of doubly occupied sites. The role of $U$ in \disp{Min-Hub} is roughly similar, at $U=0$ we have the Fermi gas, which evolves into an interacting theory with increasing $U$, giving us the Feynman-Dyson perturbation theory. The range of $\lambda$  ($\in[0,1]$) in \disp{Min-22} is bounded above, as opposed to that of $U\in[0,\infty]$ in \disp{Min-Hub}. Therefore the ECFL theory avoids dealing with a major headache of the canonical theory whenever a  coupling constant becomes large.  Recall that  realistic interactions in correlated matter  usually involve  a large coupling parameter $U$. For this purpose   one is forced to make hard-to-control approximations, such as summing specific  classes of  diagrams in different parameter ranges.  The introduction of $\lambda$ into the ECFL equations  opens the possibility  that  a low order calculation might suffice to give accurate results at low excitation energies.    This possibility is in-fact realized for important strong coupling problems as  shown earlier \refdisp{Sriram-Edward}.

 We found  in \refdisp{ECFL}  that an efficient method for proceeding with the iterative scheme is to first perform a factorization of the Greens function into two parts. The first  is an auxiliary Greens function $\GH$ satisfying a canonical equation, thus admitting a Dysonian expansion with its attendant advantage of summing a geometric series with every added term of the denominator. There remain some terms that cannot be pushed into the denominator, these are collected together as the  caparison function $\cap$. In the matrix notation used above we  first decompose the Greens function as:
  \beq
\G= \GH . \cap, \label{factor2}
\eeq
this implies a product in the $\vec{k}, \omega$ domain as written below in \disp{eq1}.
 The differential  operator $X$ in  equation \disp{Min-22} is distributed over  the two factors of \disp{factor2} using the Leibniz product rule, as 
\beq
X.\GH.\cap=\wick{\c X . \c \GH .\cap}+ \wick{\c X.  \GH. \c \cap} 
\eeq
where the contraction symbol $\wick{\c X\c a}$ indicates the term being differentiated by the functional derivative terms in $X$, while the matrix indices follow the dots.  Using $\GHI. \GH=\iden$ \disp{Min-22}  is now written as
\beq
\left(  \GHI_{0} -  \lambda {\wick{ \c X. \c \GH }}.\GHI -  \lambda {Y_1}\right). ~\GH.\cap = \delta \  ( \iden  - \lambda {\gamma}  ) + \lambda \wick{\c X. \GH . \c \cap}\; .  \label{Min-2}
\eeq
This equation  factors  exactly   into two equations upon insisting that   $\GH$ has a canonical structure:
\beq
\left(  \GHI_{0} -  \lambda {\wick{ \c X. \c \GH }}.\GHI -  \lambda {Y_1}\right) =\GHI \label{canonical}
\eeq
and 
\beq
\cap = \delta \  ( \iden  - \lambda {\gamma}  ) + \lambda \wick{\c X. \GH . \c \cap} \; .  \label{Min-3}
\eeq
We can then  use $\GH.\GHI = \iden$ to simplify the term ${\wick{ \c X. \c \GH }}.\GHI=-{\wick{ \c X. \GH. \c \GHI}} $, giving rise to a Dyson self-energy expressed in terms of a Dyson vertex functions as usual. 
The idea then is to iterate the pair of Equations (\ref{canonical},\ref{Min-3}) jointly in $\lambda$.
Details of the skeleton expansion nature can be found in \refdisp{Sriram-Edward,Edward-Sriram, ECFL}. The main point to note is that  while $\GHI,\cap$ in \disp{canonical} and \disp{Min-3} are expanded in powers of $\lambda$, the function $\GH$ is kept unexpanded as a basis term (or ``atom'') of the skeleton expansion, temporarily ignoring its relationship as the inverse of  $\GHI$.   The equal time value of the variable $\gamma$ in \disp{gamma-def} is taken from the exact sum-rule for $\G$ in \disp{number-sumrule}.
The initial values at $\lambda=0$ are $\GH= \GH_0$ and $\cap = \iden$,  and we must remember to use the product form \disp{factor2} to determine the local Greens function $\gamma$ in \disp{gamma-def}. We should note that when the source is turned off $\V\to 0$ we recover space and time translation invariance so that \disp{factor2} is simply $\G(\vec{k}, i \omega_j) = \GH(\vec{k}, i \omega_j) . \cap(\vec{k}, i \omega_j) $, with the Matsubara frequency $\omega_j = (2 j+1) \pi \beta$. At low T, the leading singularities of $\G$ are co-located with  those   of $\GH$ provided the caparison function $\cap$ is sufficiently smooth- this situation is realized in all studies done so far.

\subsection{  The shift identities and second chemical potential $u_0$  \label{second-chem} }
Before proceeding with the iterative scheme, it is important to discuss a simple but crucial symmetry of the \tJ model- {\em the shift invariance}, first noted   in \refdisp{Monster}.  In an exact treatment shifting  $t_{ij} \to t_{ij}+ c_t \; \delta_{ij}$ with    $c_t$ arbitrary, is easily seen to be innocuous, it   merely adds  to   \disp{hamiltonian}  a term $-c_t \sum_{\si}\hat{N}_\si$ whereby the center of gravity of the band is displaced.  (Here $\hat{N}_\si$ is the number operator for electrons with spin $\si$.) However in    situations such as the $\lambda$ expansion,  the Gutzwiller constraint is released at intermediate values, here  it has  the  effect  of adding terms derivable from a local (i.e. Hubbard type)   interaction  term. \footnote{Similarly we note that shifting $J_{ij}\to J_{ij} + c_J \delta_{ij}$ with arbitrary $c_J$ also adds a similar unphysical local interaction term, as discussed in greater detail in \refdisp{Monster}} To see this consider the fundamental commutator term $[H_{tJ},\X{i}{0\si_i}]$ in
\disp{commutator},  here under the shift $t_{ij} \to t_{ij}+ c_t \; \delta_{ij}$, the third term gives rise to an extra term $c_t \X{i}{\sib_i \sib_i} \X{i}{0 \si_i}$. This term vanishes only in a Gutzwiller projected state,  the equations of motion by themselves do not eliminate it. 
  Its appearance is tantamount to adding a Hubbard like term $\frac{c_t}{2} \sum_i \X{i}{\si \si} \X{i}{\sib \sib}$ to the Hamiltonian $H_{tJ}$.   As argued in \refdisp{Monster} we would like the equations of motion for the Greens functions to be explicitly  invariant under the above shift of $t_{ij}$ to {\em each order in $\lambda$}. 
Enforcing this shift invariance  to each order in  the $\lambda$ expansion plays an important ``watchdog'' role on the $\lambda$ expansion.
  
   An efficient method to do so is to explicitly introduce an extra Lagrange multiplier $u_0$ through   a term
$\lambda u_0 \sum_i N_{i \up} N_{i \dn}$ to the Hamiltonian \disp{hamiltonian}. This amounts to replacing $t_{ij}\to t_{ij} + \delta_{ij} \frac{u_0}{2}$ in all terms other than in the bare propagator $\GH_0$. The 
    $u_0$ term  makes no difference when  $\lambda$ is set at unity in the exact series, since  double occupancy is excluded. In practice, we set $\lambda=1$ in equations that are truncated at various orders of $\lambda$, and the magnitude of $u_0$ is fixed through a second constraint. We thus  have two variables to fix, namely $u_0$ and $\chem$. We also have two constraints, the number sum-rules  $n_{\si}= \GH_{\si \si}(i \tau^-, i \tau)$, and  $n_{\si}= \G_{\si \si}(i \tau^-, i \tau) ($\disp{number-sumrule}). In the absence of a magnetic field the number densities $n_{\si}$ reduce as $n_{\si} \to \frac{n}{2}$, where $n$ is the number of particles per site.

  After turning off the sources, in the momentum-frequency space we can further  introducing two self energies $\Psi(k, i\omega_j)$, and $\Phi(k, i\omega_j)$
with
\beq
\cap(\vec{k}, i \omega_j)&=& 1- \lambda \frac{n}{2} + \lambda \Psi(\vec{k}, i \omega_j) \label{caparison} \\
\GH^{-1}(\vec{k}, i \omega_j)&=& \GH_{0}^{(-1)}(\vec{k}, i\omega_j)  + \lambda \left(
\frac{n}{2} \varepsilon_k +  \frac{n}{4} J_0\right)
  - \lambda \Phi(\vec{k}, i \omega_j). \label{auxg}
\eeq
Here $\varepsilon_k$ and $J_k$ are the Fourier transforms of $-t_{ij}$ and $J_{ij}$.
In the right hand side of \disp{auxg},  the second and third terms arise respectively  from the equal-time limit of $\lambda Y_1$ 
and  $ \lambda \wick{ \c X. \c \GH }.\GHI$ in \disp{canonical} respectively.  The two self energies   $\Phi, \Psi$  are explicitly $\lambda$ dependent, they vanish at  infinite frequency for any $\lambda$.  Thus we write 
\beq 
\G(k, i \omega_j)=
  \GH(k, i \omega_j) \times \widetilde{\mu}(k, i \omega_j).  \label{eq1} 
\eeq
The auxiliary Greens function satisfies a second sum-rule that is identical   to \disp{number-sumrule}, both may written in the Fourier domain: 
\beq
(k_B T) \sum_{k, j} e^{ i \omega_j 0^+}  
   G_{\si \si}(k, i \omega_j)    
 = n_\si  \label{second-sumrule}; \; \mbox{for both}\;  G= \G \; \mbox{and} \; \GH.
\eeq
 \disp{factor2} can now be written explicitly  in the non-Dysonian form proposed in \refdisp{ECFL}
 \beq
\G(\vec{k}, i\omega_j)= \frac{1- \lambda \frac{n}{2} + \lambda \Psi(\vec{k}, i\omega_j)}{\GH_{0}^{(-1)}(\vec{k}, i\omega_j)  + \lambda  \frac{n}{2} \varepsilon_k + \lambda \frac{n}{4} J_0 - \lambda \Phi(\vec{k}, i\omega_j)}. \label{twin-self}
\eeq
We observe that simple Fermi liquid type self energies $\Psi$ and $\Phi$ can, in the combination above,   lead to highly asymmetric (in frequency) Dyson self energies \cite{ECFL,Monster,Sriram-Edward,Edward-Sriram}. Finally we note that our calculations are performed in terms of spectral function obtainable from analytic continuation of the Matsubara frequencies into the upper complex half plane of frequencies:
\beq
\rho_{\G}(\vec{k}, \omega)  & = & -\frac{1}{\pi} \; \Im m \,\G(\vec{k}, i \omega_j \to \omega+ i 0^+), \nn \\
\G(\vec{k}, i \omega_j) &=& \int \frac{\rho_{\G}( \vec{k}, \omega)}{i \omega_j - \omega} \label{spectral-functions}, 
\eeq
and similarly defined spectral functions for variables $\GH,\Phi,\Psi$ etc.

\subsection{\bf Summary of equations to second order in $\lambda$  \label{secondorderequations}} 
In the following, we use the minimal second order equations \cite{Sriram-Edward,Edward-Sriram,WXD} obtained by expanding  \disp{canonical} and \disp{Min-3} to second order in $\lambda$. The calculation is straightforward and a systematic notation is detailed in \refdisp{Edward-Sriram}, which is followed here. We use the abbreviation\cite{notation-1} $k \equiv (\vec{k}, i\omega_n)$, and  also redefine $\Phi(k) = \chi(k) + \varepsilon_k \Psi(k)$,  keeping in mind that one set of terms in $\Phi$ have an external common factor of $\varepsilon_k$ multiplied by all terms in $\Psi$. We next collect the answers below in terms of the two self energies $\chi,\Psi$  
 \beq 
 \GH^{-1}(k)=i\omega_n+{ \chem} -   \underbrace {\varepsilon_k} + \lambda \frac{1}{4}n J_0  -   \varepsilon_k (- \lambda \frac{n}{2} + \lambda \Psi)-  \lambda \chi(k). \label{eq22}
 \eeq
 We  now expand $\Psi$ and $\chi$ from \disp{canonical} and \disp{Min-3} in powers of $\lambda$.  To the lowest two orders we find  $\Psi= \lambda \Psi_{[1]} +O(\lambda^2)$ and $\chi= \chi_{[0]} + \lambda \chi_{[1]}+O(\lambda^2)$, where  $\chi_{[0]}= - \sum_p \GH_p (\varepsilon_p + \frac{1}{2} J_{k-p})$.

The next step is to  introduce $u_0$ explicitly: we write $\varepsilon_k\to \varepsilon'_k= \varepsilon_k -\frac{u_0}{2} $
in every occurrence of $\varepsilon_k$, except in the bare propagation term (the term with an underbrace) in  \disp{eq22}.   
 \beq 
 \GH^{-1}(k)=i\omega_n+{ \chem} + \lambda \frac{1}{4}n J_0 - \frac{1}{2}u_0 - \widetilde{\mu}(k)  \varepsilon'_k-  \lambda \chi_{[0]}(k) -  \lambda^2 \chi_{[1]}(k). \label{eq2}
 \eeq
 Note that  the shift with $u_0$ also applies to the term   $\chi_{[0]}$, it now reads  $\chi_{[0]}= - \sum_p \GH_p (\varepsilon'_p + \frac{1}{2} J_{k-p}) $.  We note the  expressions for $\chi_{[1]}, \Psi_{[1]} $ from \refdisp{Edward-Sriram} Eq.~(65-67):
 \beq
 &&\chi_{[1]}(k)=
 -  \sum_{pq} \left(\varepsilon'_p+\varepsilon'_q +\frac{1}{2}(J_{k-p}+J_{k-q})\right)\times(\varepsilon'_{p+q-k}+  J_{q-k}) \GH(p)\GH(q)\GH(p+q-k), \nn \\
  \label{eq4}
\eeq
\beq
 \Psi_{[1]}(k)  = -   \sum_{pq}(\varepsilon'_p+\varepsilon'_q+J_{k-p})\GH(p)\GH(q)\GH(p+q-k),  \label{eq3}
\eeq

  We now  set $\lambda=1$ and record the final equations:
 \beq
 \GH^{-1}(k)&=&i\omega_n+ \left( \chem +  \frac{1}{4}n (J_0-u_0) - \frac{1}{2}u_0+ \sum_p \GH_p \varepsilon_p+  \frac{J_k}{2} \sum_p \GH_p \cos\, p_x \right) - \widetilde{\mu}(k)  \varepsilon'_k-   \chi_{[1]}(k), \nn \\ \label{eq40} \\
 \cap({k})&=& 1-  \frac{n}{2} +  \Psi_{[1]}({k}), 
 \label{eq23}
 \eeq
where we used a nearest neighbor $J_{ij}$ and cubic symmetry in the simplifications. 
We can verify that the above expressions obey the  shift invariance: if we shift $\varepsilon_k \to \varepsilon_k +c_0$, the arbitrary constant $c_0$ can be absorbed by shifting $\chem \to \chem+ c_0$ and $u_0\to u_0+ 2 c_0$,  and is thus  immaterial. 
The band energy is given explicitly  as $\varepsilon_k= - 2t (\cos(k_x a_0)+ \cos(k_ya_0)) - 4 t' \cos(k_x a_0) \cos(k_y a_0)$, where t and $t'$ are the first and second neighbor hopping amplitudes. 

\subsection{High energy cutoff scheme \label{cutoff}}

The self consistent  solution of the second order equations of  Eqs.~(\ref{eq4},\ref{eq3},\ref{eq40},\ref{eq23}) plus the number sum-rules, can be found numerically by discretizing the momentum and frequency variables on a suitable grid.  This procedure   can be carried out in a straightforward way for low $T\lessim t$ and high hole densities $\delta \gssim 0.3$ (low particle densities $n \lessim 0.7$). At lower hole densities or at high temperature $T \gg t$, the equations run into convergence problems. The origin of this problem is  the formation of weak and featureless tails of the spectral functions  extending to quite high energies. These tails are known to be artificial, since they do not occur in the exact numerical solutions where available. Thus the second order theory seems  insufficient  in the regime of low hole densities $\delta \lessim 0.2$, where much of the current interest lies. A  technically rigorous  resolution of the problem of weak tails seems  possible. However it requires the  non-trivial calculation of higher order terms in the $\lambda$ expansion. Such higher order  terms oscillate in sign and hence cancellations at high energies are expected. 

In view of the substantial  magnitude of the program of summing the $\lambda$ series to high orders, it seems worthwhile to investigate simpler and  physically motivated approximations for improving the lowest order scheme.  It turns out that there are  a few interesting alternatives in this direction.  In \refdisp{Sriram-Edward}  we showed one convenient way  to handle the high energy  tail problem practically, through the introduction of a high energy cutoff. The choice of an objective cutoff  
was rationalized by considering two physically different limits, that of high particle density $n\to1$ and the simpler high temperature limit, where related tails are found. The cutoff is  chosen  using the analytically available high T limit results and  then applied to all densities  and T. 

The cutoff scheme of \refdisp{Sriram-Edward} is not rigorous, but enables us to extract meaningful results for low energy excitations  from the second order $\lambda$ equations, out to  fairly  low hole densities $\delta \lessim 0.2$.    It is benchmarked  in the case of $d=\infty$, where the cutoff scheme quantitatively  reproduces the spectral weights in the most interesting regime of low energies $|\omega| \ll t$, while erring somewhat at energies above the scale of quarter bandwidth. In \refdisp{Sriram-Edward,WXD} the resulting physical quantities such as resistivity are shown to be in good correspondence to the exact results from DMFT. In view of this success we use a similar  cutoff scheme for 2-d below, with the expectation that the physics of the low energy excitations is captured.  In the present 2-d case we employ a single (re)-normalization the spectral function for each $\vec{k}$ as
\beq
\hat{\rho}_{\GH}(\vec{k},\omega) = \frac{1}{{\cal N}_k} W_T(\omega- \varepsilon_k) \rho_{\GH}(\vec{k},\omega), \label{Tukey}
\eeq
where $W_T$ is a smooth window (even) function shown in Fig.~(3) \refdisp{Sriram-Edward}. It is centered on the bare band energy and has  width $4D$, where $2 D$ is  the bandwidth ($\sim 8 t$ in this case).  The  constant ${\cal N}_k$ is fixed by the normalization condition $\int \hat{\rho}_{\GH}(\vec{k},\omega) d \omega =1$.  In the present case of 2-d we can impose this cutoff window at each $\vec{k}$ individually, so that only $\vec{k}$ states very far from the chemical potential are affected.

  The two chemical potentials $\chem$ and $ u_0$ are determined through the number sum rules written in terms of the Fermi function $f(\omega)=(1+e^{\beta \omega})^{-1}$ and the spectral functions:
  \beq
\sum_{k} \int \hat{\rho}_{\GH}(k,\omega) f(\omega) d \omega = \frac{n}{2}=  \sum_{k} \int \rho_{\G}(k,\omega) f(\omega) d \omega . \label{sumrule}
 \eeq
 The set of  equations Eqs.~(\ref{eq4},\ref{eq3},\ref{eq40},\ref{eq23}, \ref{Tukey}, \ref{sumrule}) constitute the final set of equations to be computed. These are valid   in any dimension, and    reduce  to the ones benchmarked  in $d=\infty$ after setting  $J\to0$\cite{Sriram-Edward,WXD}.

  After analytically continuing $i\omega_n\to \omega+ i 0^+$ we determine the  spectral function of the interacting electron spectral function $\rho_{\G}(\vec{k},\omega)= -\frac{1}{\pi}  \Im m \, \G(\vec{k},\omega)$. The set of Equations~(1-5)  was solved iteratively on $L\times L$ lattices with $L=19,37,61$ and  a frequency grid with $N_\omega=2^{14},2^{16}$ points.  Other details are essentially the same as in our recent study of the $d=\infty, J\to0$ case  in  \refdisp{Sriram-Edward,WXD}.

\section{Results}

\S {\bf Band Parameters:}
 The \tJ model   is studied on the square lattice  with hopping parameters $t$  and $t'$ for first and second neighbors. The hopping parameter $t>0$, while  $t'$ is varied between $-0.4 t$ and $0.4 t$, thereby  changing  the  Fermi surface (FS)  from hole-like to  electron-like.  Parameters  relevant to  cuprate High Tc materials are summarized in\cite{tJ-review1,exptl-tprime,Bansil}.   Following \cite{tJ-review1}   we assume   $t\sim 0.45$ eV, giving a bandwidth  $\sim 3.6$ eV.

\begin{figure}
 \includegraphics[width=.5\columnwidth]{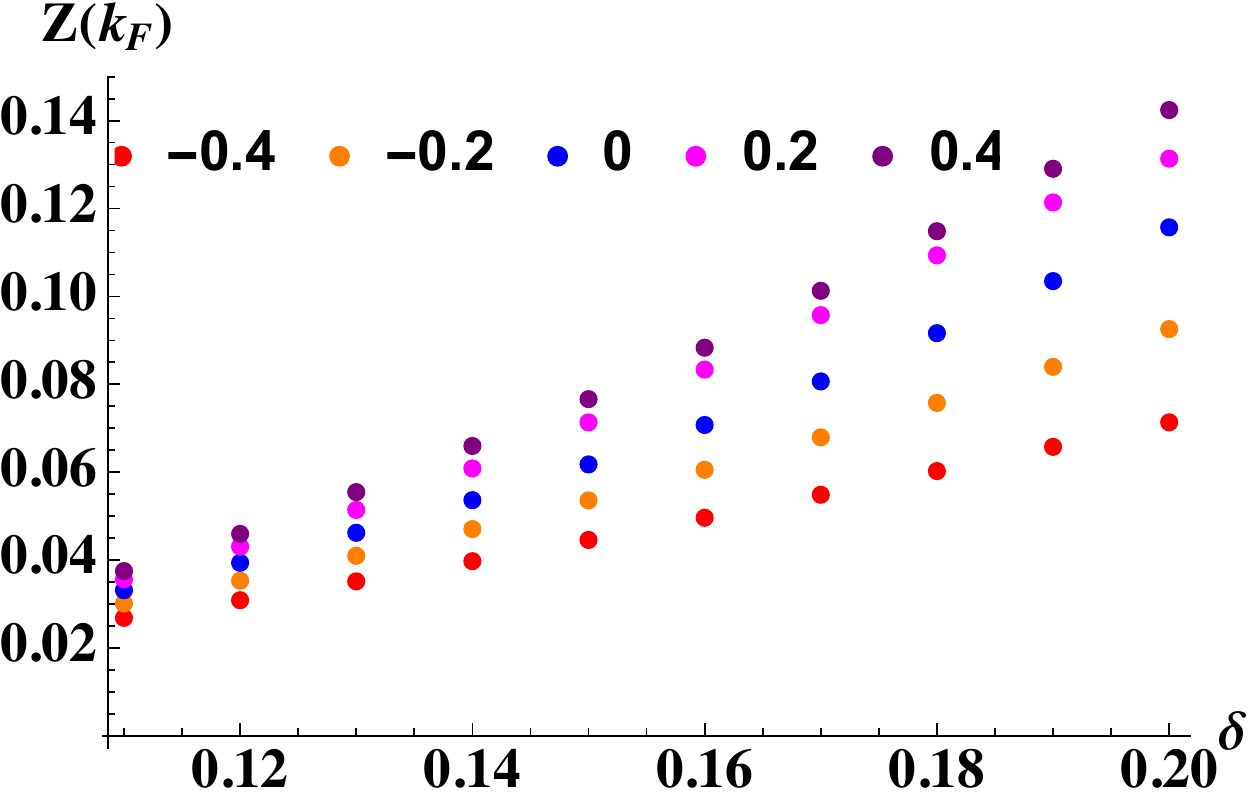}
 \caption{Hole density $\delta$, and $t'/t$ variation of  the nodal $Z(k_F)$ at $T=63$K.  $t'/t$  is  marked at the top.  Decreasing $t'$  has a similar effect to decreasing $\delta$.     \label{Fig1} }
 \end{figure}

 \S {\bf Single-particle spectrum:}
The quasiparticle energy $E(\vec{k})$ and quasiparticle weight $Z(\vec{k})$ are found from $\G$ as usual\cite{Sriram-Edward}.
In \figdisp{Fig1} we display the hole density $\delta$  and $t'$ dependence of the low temperature  $Z(k_F)$,   along the nodal (i.e. $\langle 11 \rangle$) direction.  The typical  magnitudes of $Z$ are comparable or lower than   those reported in $d=\infty$\cite{Sriram-Edward}.  A {\em new and  important  feature} is the    strong sensitivity of $Z(k_F)$  to the sign and magnitude of $t'/t$. Both  decreasing $t'$ (at fixed $\delta$) and decreasing $\delta$  (at fixed $t'$)  reduce $Z$. This feature is basic to understanding our main results. 
{We  next study   the  decay rate of the electrons 
\beq
\Gamma(\vec{k})=  -Z(\vec{k})\times   \, \Im m \, \Sigma(\vec{k},E(\vec{k})), \label{decayrate}
\eeq
 found as the half-width at half-maximum of the spectral function $\rho_{\G}(\vec{k},\omega)$  at fixed $\vec{k}$.  We display the T variation of  $\Gamma$ and $-\Im m \, \Sigma$ at the Fermi surface for three representative values of $t'/t$ in \figdisp{Fig2}. Both variables display considerable variation
with    modest change of $T$. The case of  $t'>0$ shows a distinct quadratic T dependence, but for $t'\leq 0$ we note the strong reduction, or absence, of such a quadratic dependence. Below we note a closely parallel  T and $t'$ dependence of the resistivity. 
 \begin{figure}
 \includegraphics[width=.5\columnwidth]{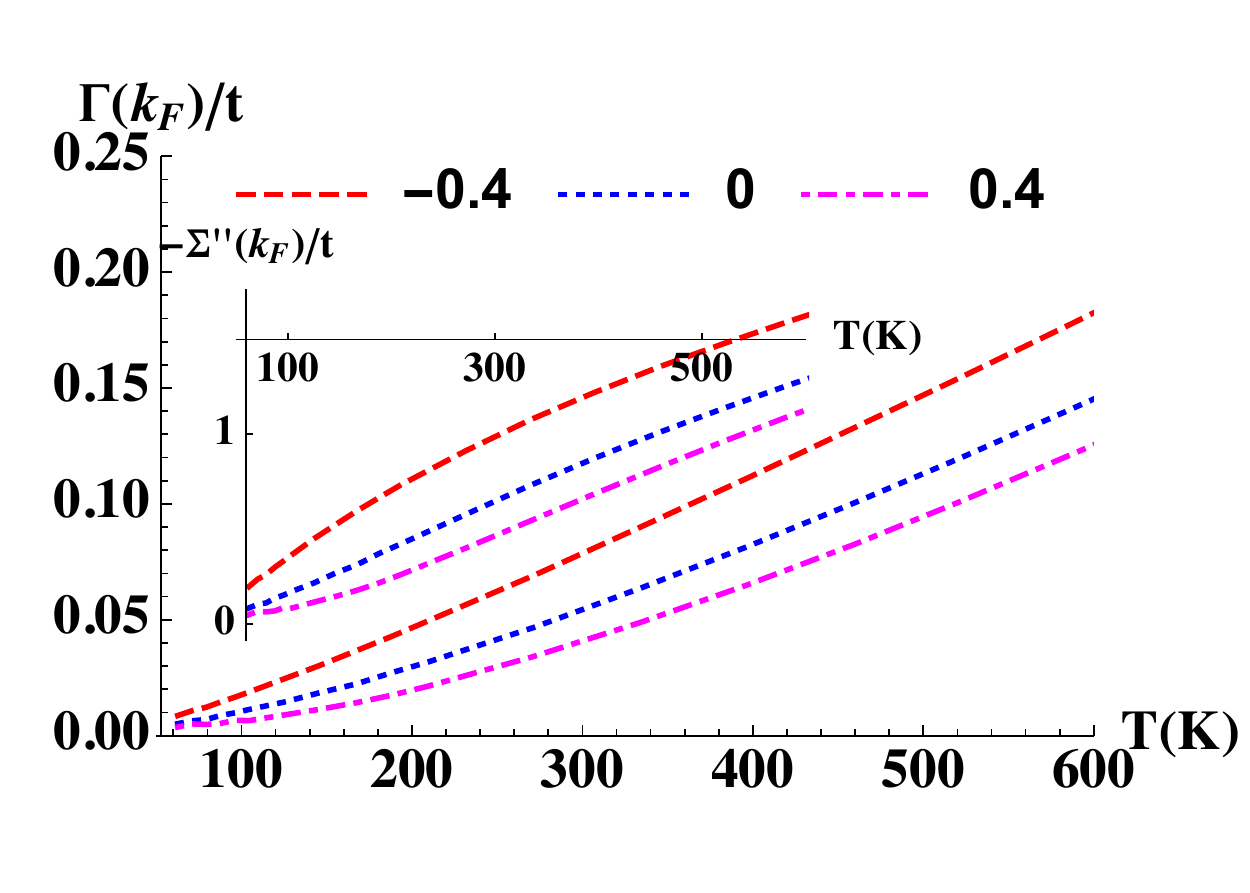}
 \caption{ $\delta=0.15$:  The decay rate \disp{decayrate}  and (inset) the nodal  $- \Im m \, \Sigma(k_F)$.  $t'/t$ is  marked at top. While $t'=0.4 t$ has a positive curvature for both variables, $t'=-0.4t$ displays a prominent negative curvature in $\Sigma''$ (i.e. $\Im m  \Sigma$), and an almost flat $\Gamma$.
  \label{Fig2} }
 \end{figure}
 }

\begin{figure}
\subfigure[  \;\; t'/t=-0.4, T= 63K]{\includegraphics[width=.3\columnwidth]{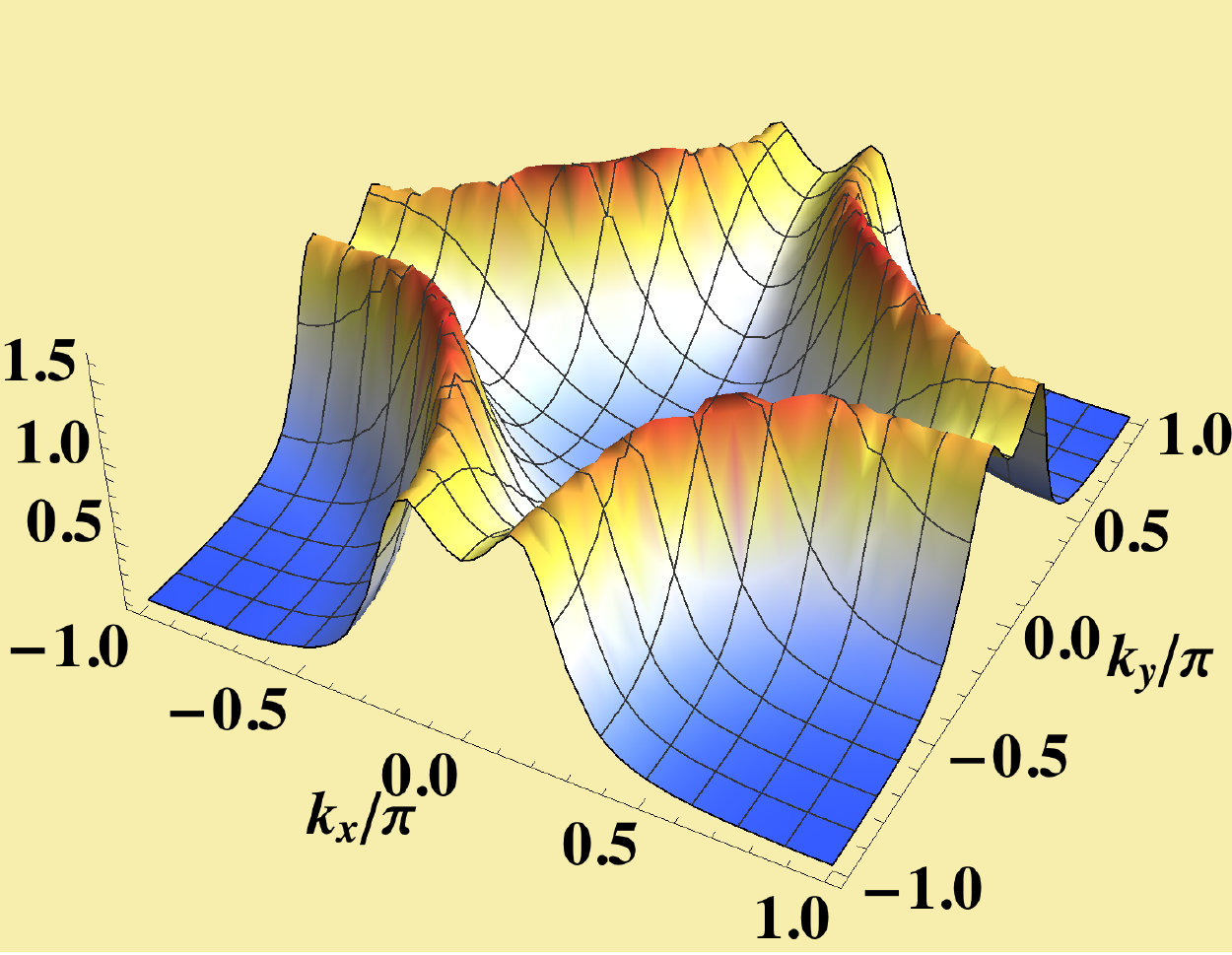}}
\subfigure[\;\; t'/t=-0.4, T= 210K]{\includegraphics[width=.3\columnwidth]{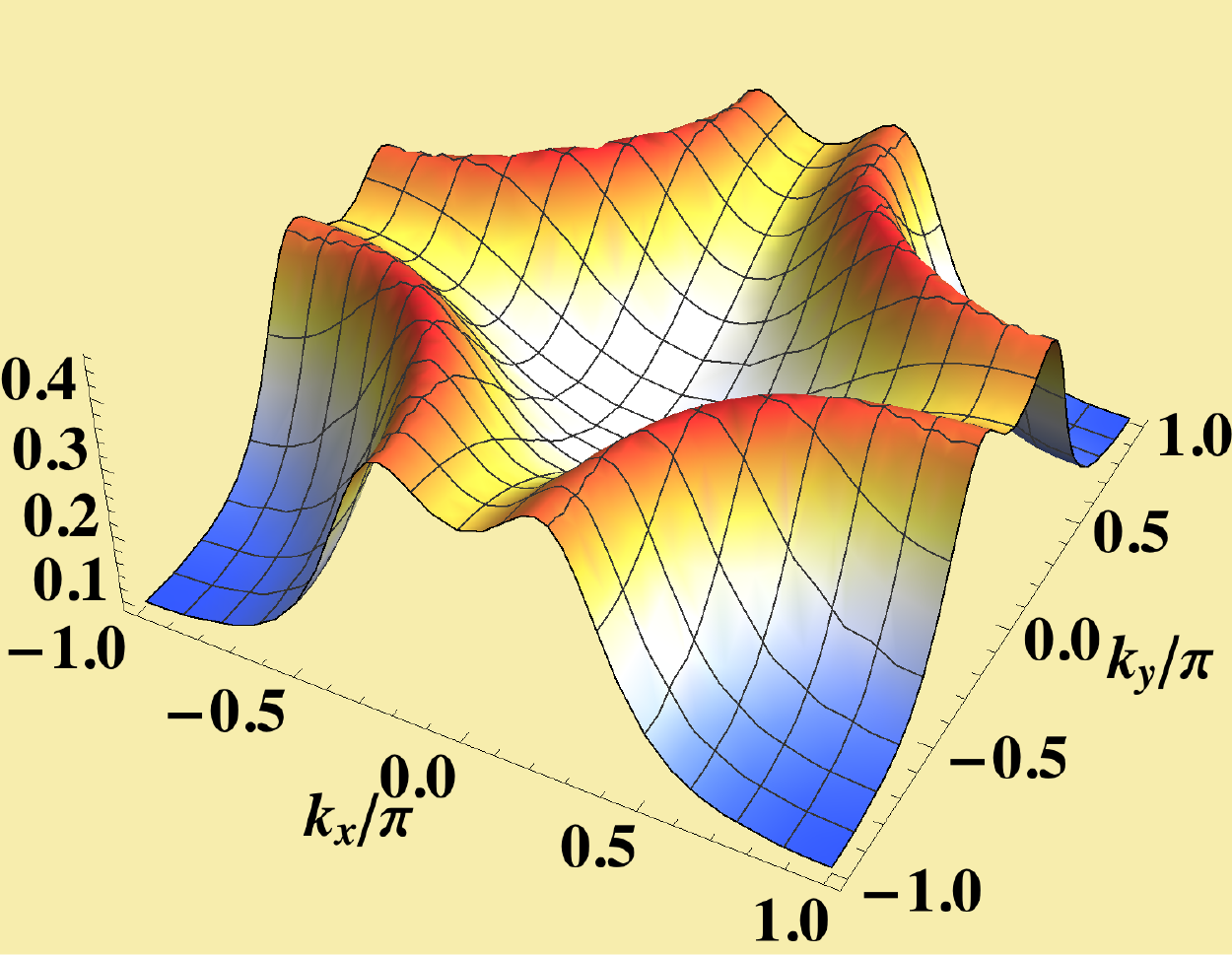}}
\subfigure[\;\; t'/t=-0.4, T= 334K]{\includegraphics[width=.3\columnwidth]{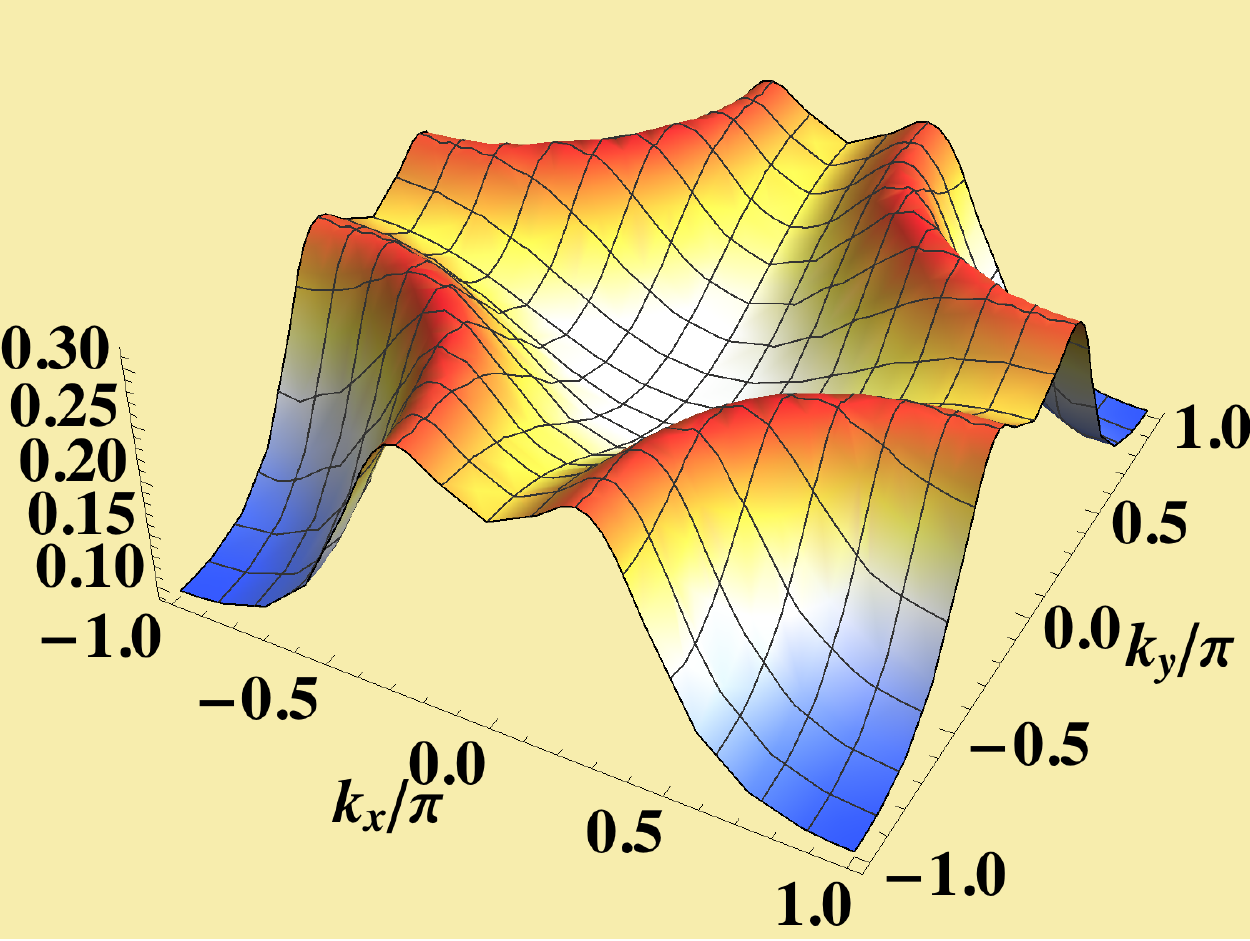}}
\subfigure[ \;\; t'/t=0, T= 63K]{\includegraphics[width=.3\columnwidth]{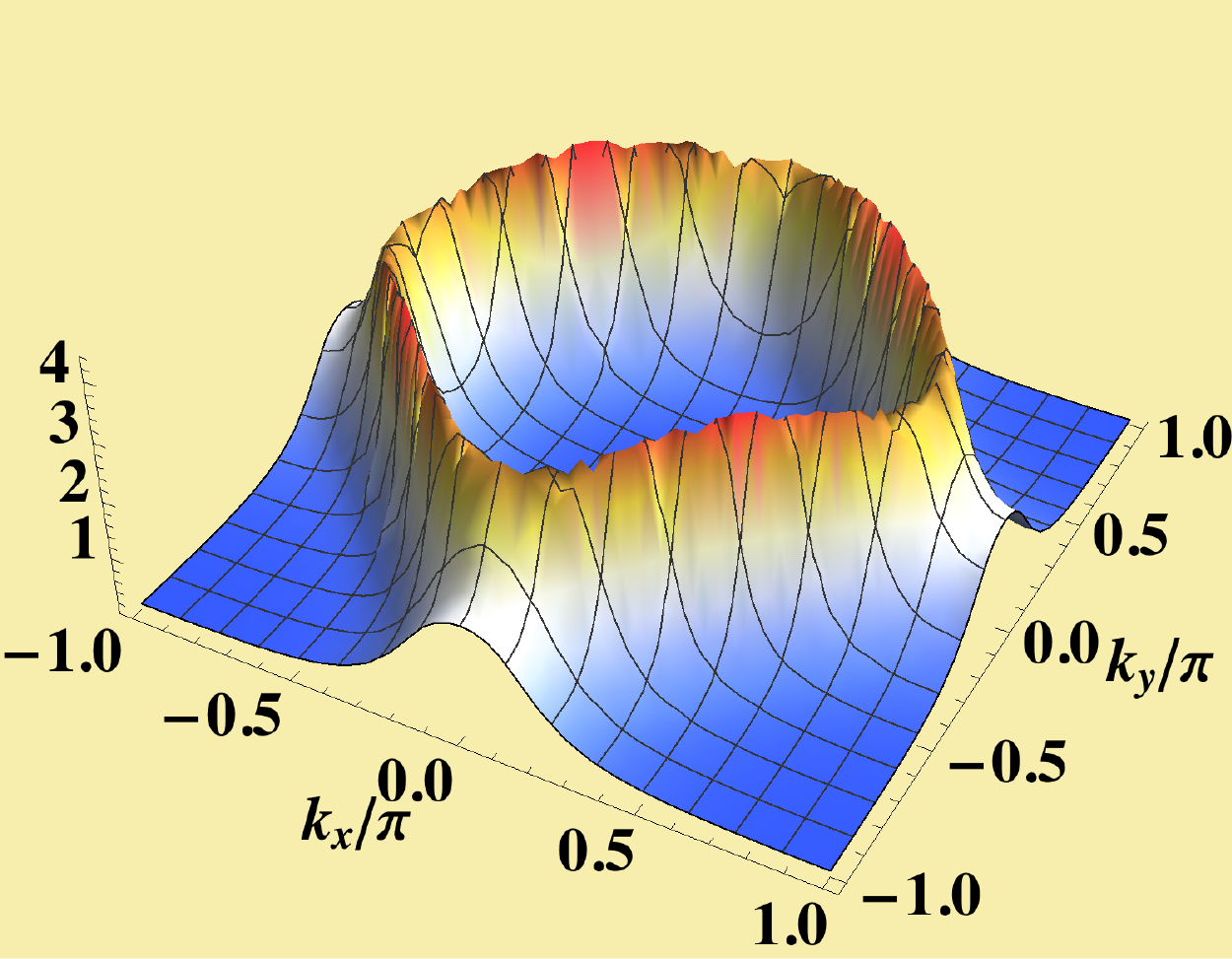}}
\subfigure[\;\; t'/t=0, T= 210K]{\includegraphics[width=.3\columnwidth]{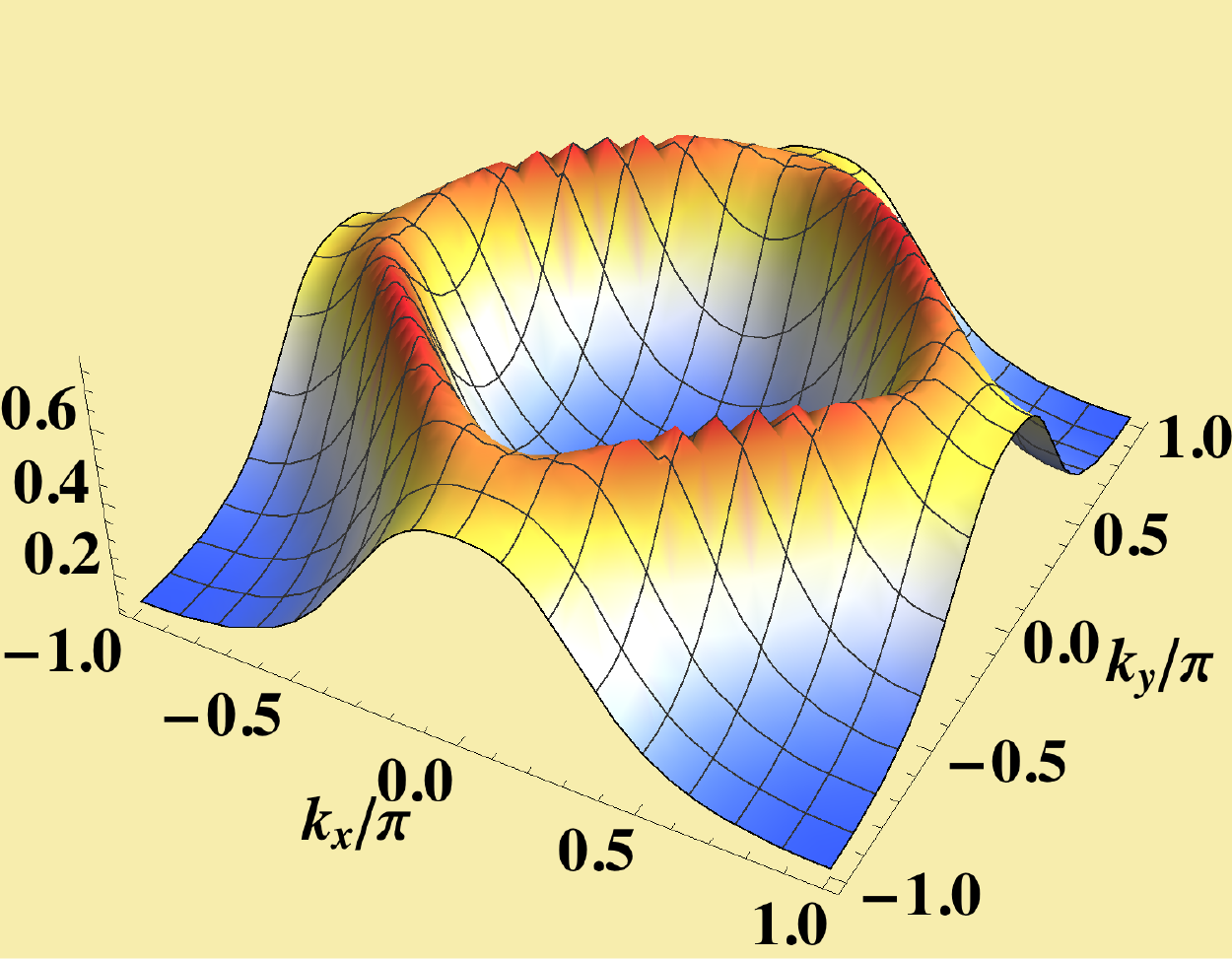}}
\subfigure[\;\; t'/t=0, T=334K]{\includegraphics[width=.3\columnwidth]{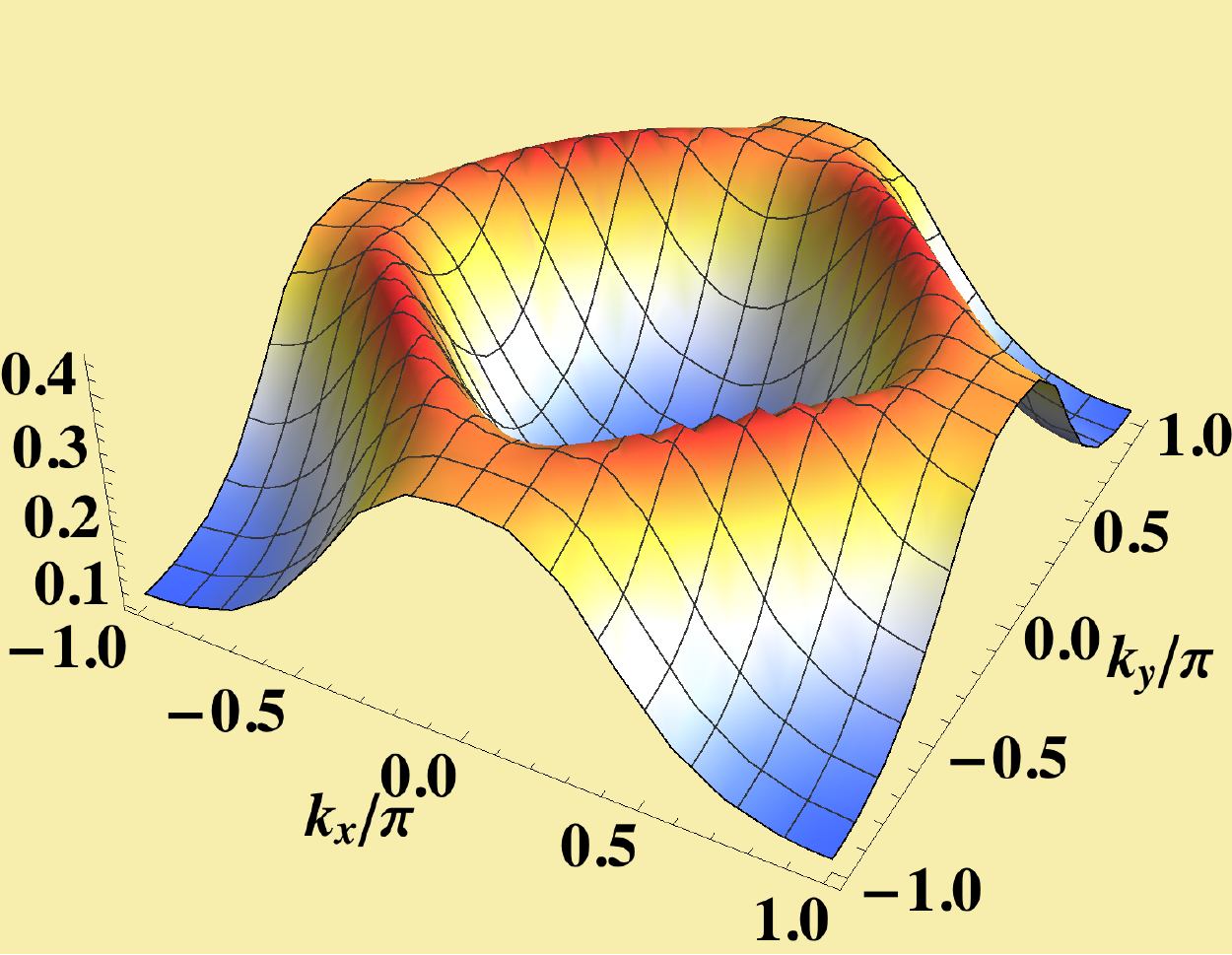}}
\subfigure[  \;\;t'/t=0.4, T= 63K]{\includegraphics[width=.3\columnwidth]{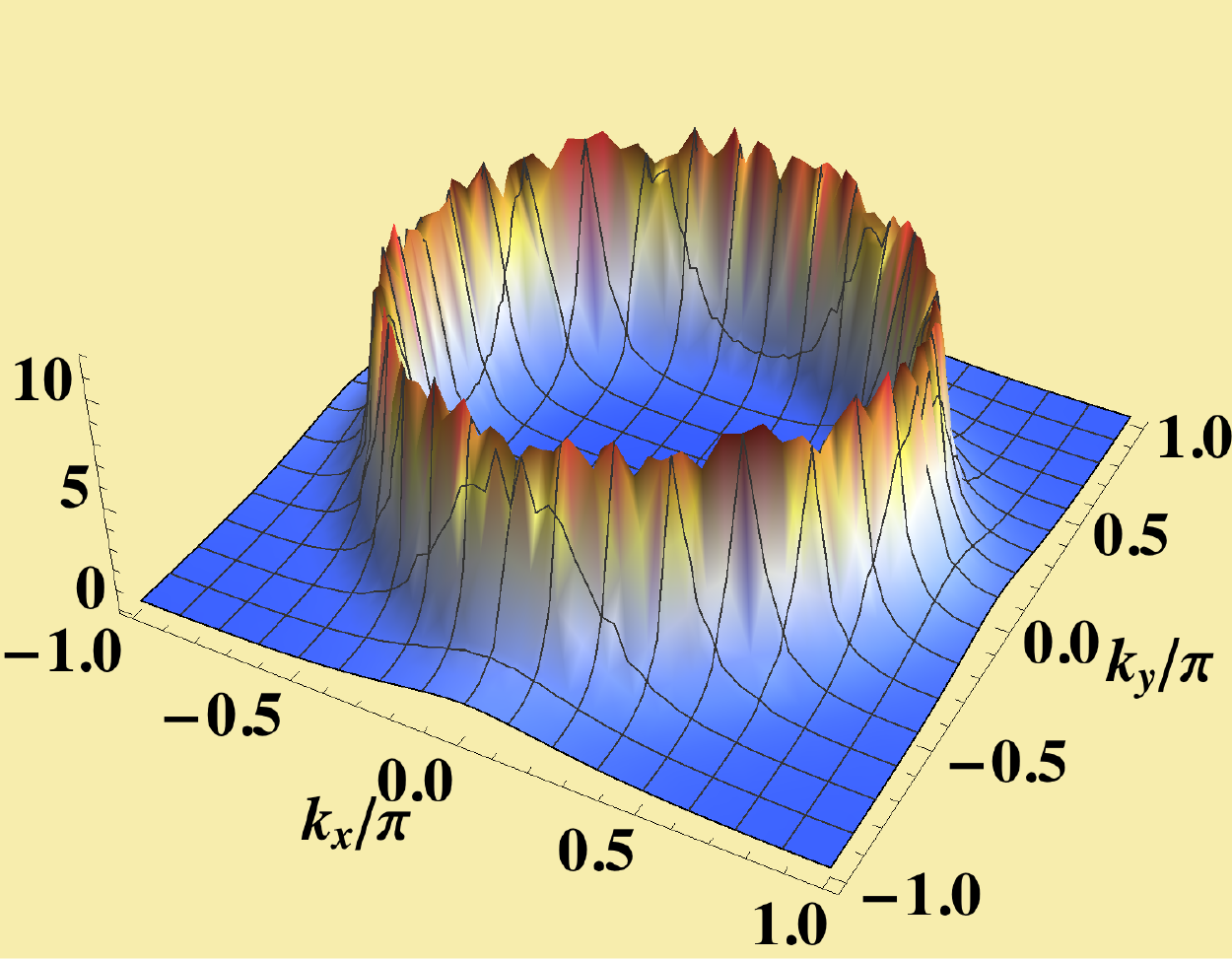}}
\subfigure[\;\; t'/t=0.4, T= 210K]{\includegraphics[width=.3\columnwidth]{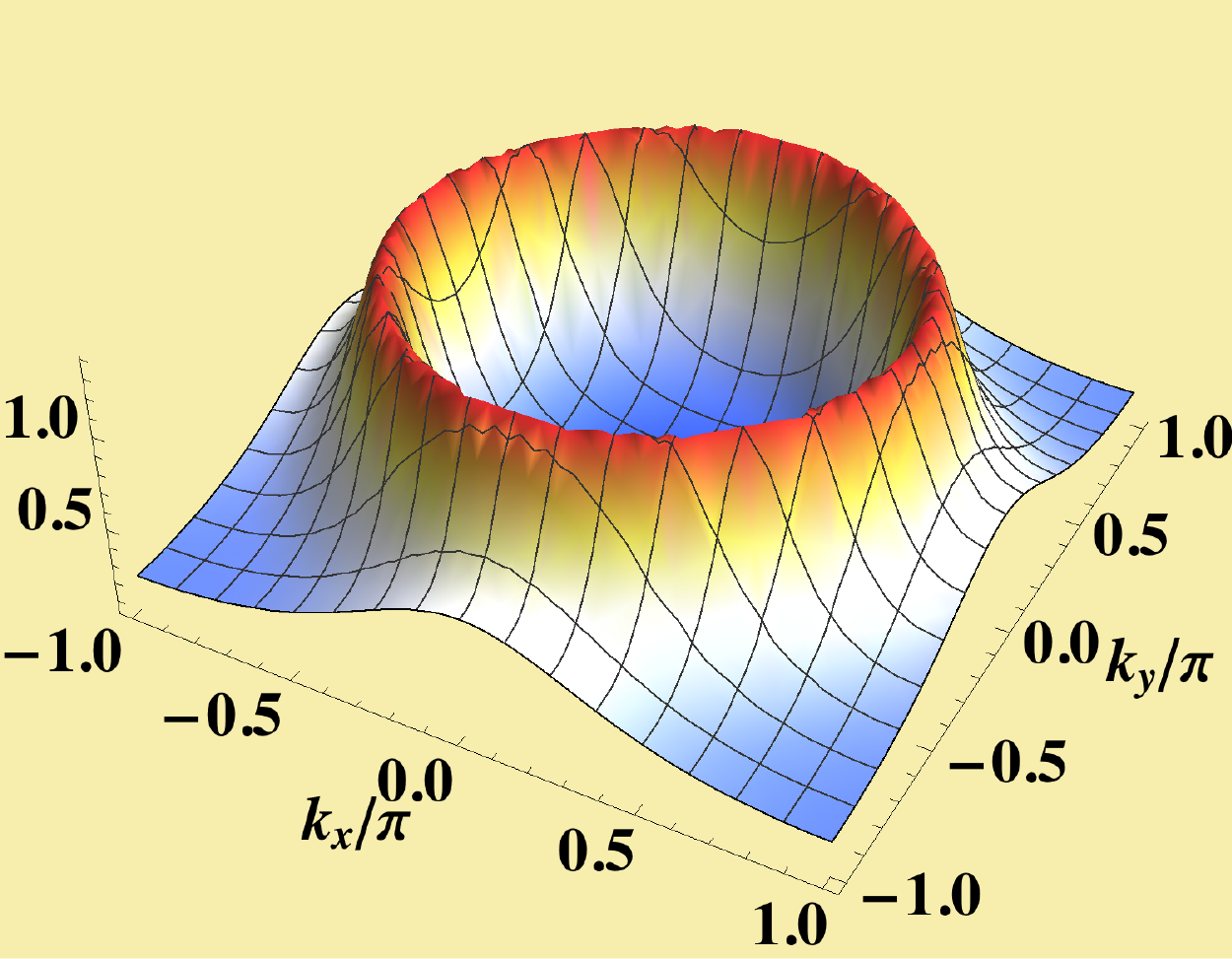}}
\subfigure[\;\;t'/t=0.4, T= 334K]{\includegraphics[width=.3\columnwidth]{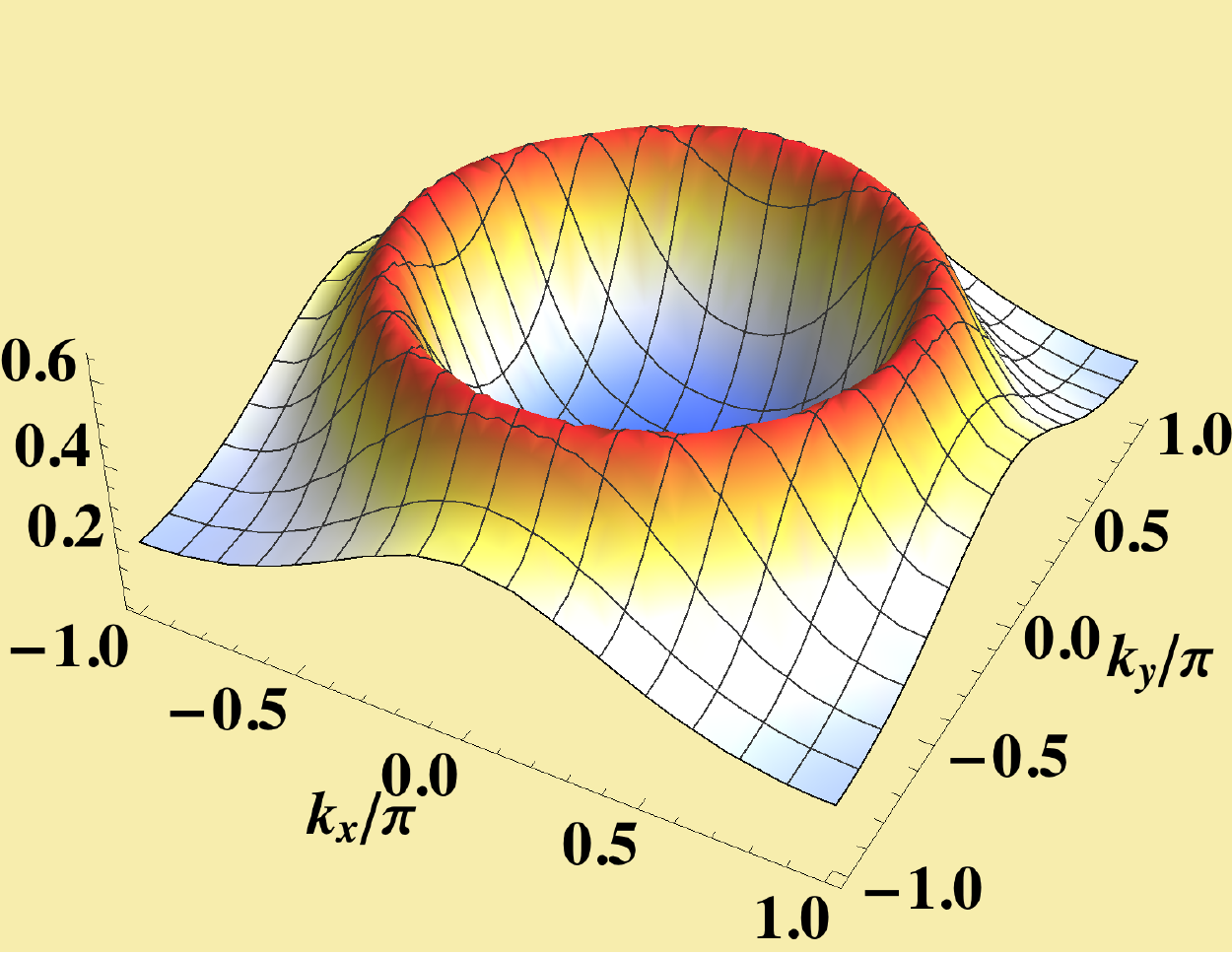}}
\caption{$\delta=0.15$.  The peak height of the spectral function $A(\vec{k},\omega)$ over the Brillouin zone at various  $t'$ and $T$. At low temperatures the steep increase  of the peak heights in going from Panels $(a)\to (d) \to (g) $ illustrates the almost FL nature of $(g)$ 
$t'>0$, relative to $(a)$ with $t'<0$. The complementary view of variation with $T$ in going from Panels $(a)\to (b) \to (c) $ etc illustrates the dramatic thermal sensitivity in all cases. Recalling that our bandwidth is $\sim3.6$ eV,  we observe that  the tiny $0.35\%$ variation of temperature relative to the bandwidth,  in warming  from 63K  to  210K  drops the peak height by a factor between $5$ and  $10$, followed by a more gentle fall to 334K. This extraordinary thermal sensitivity is characteristic of our solution of the \tJ model, it is also reflected in other variables discussed here such as the resistivity. 
 \label{Fig3} }
\end{figure}

 In \figdisp{Fig3} we display the photoemission accessible  peak heights of the spectral function  $ \{t\star \rho_{\G}(\vec{k},\omega)\}_{max}  $ over the BZ at three representative values of $t'/t$, at three temperatures $T=63,210,334$K.  The peaks track the  non-interacting FS,   changing from hole-like in Panels (a,b,c) to strongly  electron-like in Panels (g,h,i).   Several  features are noteworthy. The peaks are higher in the nodal  relative to the anti-nodal direction at low T.   We observe the high sensitivity to warming, in going from $T=63$K to $T=334$K  a small ($\sim0.7\% $)  change in $T$ relative to the bandwidth  causes a five to fifteen-fold  drop in the spectral  peaks at the Fermi surface. This is correlated to the thermal variation of $-\Im m \, \Sigma$ at the same set of $t'$, shown in the inset of \figdisp{Fig2}, since the intensity at $k_F$ is essentially the inverse of this object.  Meanwhile  the background spectral weight rises rapidly in all cases, to a roughly similar  magnitude. The figure  shows that at low $T$ the curve with $t'>0$ has much higher peaks than $t'\leq 0$, giving the impression of weaker correlations. However  the drop on warming  is   the  largest in this case, which  signifies another facet of strong correlations.  {\color{blue} }

 \S {\bf Resistivity:} We now study the behavior of the resistivity from electron-electron scattering. We  use the popular bubble approximation,   factoring the current correlator as $\langle J(t) J(0) \rangle \sim  \sum_k v_k^2 \G^2(k)$, where the bare current vertex is the velocity 
$\hbar v_k^\alpha=  \frac{\partial \varepsilon_k}{ \partial k_\alpha}$. In tight binding theory $v_k^\alpha$   oscillates in sign, resulting in  a vanishing average over the Brillouin zone.  This oscillation is expected to reduce magnitude of the vertex corrections \cite{bubble}.
 For a  3-d   metal having  well separated sheets  in the c  direction  ($c_0$ the separation of the sheets), with each sheet represented  by the  2-d \tJ model,    the DC resistivity $\rho_{xx}$  can  be written in terms of the electron spectral function as follows. 
We define  a dimensionless resistivity $\bar{\rho}_{xx}$ whose inverse is given by
 \beq
  \bar{\sigma}_{xx} =   \langle\Upsilon(\vec{k}) ({\hbar v_k^x})^2/{a^2_0} \rangle_k, \label{parallel}
 \eeq
where $\langle A \rangle_k \equiv  \frac{1}{N_s}\sum_{\vec{k}} A(\vec{k}) $, while   the momentum resolved   relaxation scale is:
\beq
\Upsilon(\vec{k}) =(2 \pi)^2 \int_{-\infty}^\infty d\omega \, (- {\partial f}/{\partial \omega}) \rho^2_G(\vec{k},\omega), \label{upsilon}
\eeq
and  $f\equiv 1/(1+\exp{\beta \omega})$ is the Fermi function. This object resembles the spectral peaks in \figdisp{Fig3}, losing height and broadening rapidly with T. The physical 3-d resistivity is given by
$\rho_{xx}= \rho_0 \times \bar{\rho}_{xx}$, where  $\rho_0\equiv c_0 h/e^2 $ ($\sim1.71$m$\Omega$ cm) serves as    the scale of resistivity\cite{lattice}, and using the measured  values of the lattice constants  we can express our results in absolute units. For understanding the magnitude  of the inelastic scattering it can be useful to convert the resistivity into the dimensionless parameter $ \langle  k_F  \rangle \, \lambda_{m}$ of an effective 2-d continuum theory, where $\lambda_{m}$ is the  mean-free-path and where $\langle k_F \rangle$ is an   (angle averaged) effective Fermi momentum.  We can use a relation argued for in \refdisp{Ando2,Greg2}
\beq
\langle k_F \rangle \;  \lambda_{m} = \frac{h c_0}{e^2 \rho_{xx}}= \frac{1}{\bar{\rho}_{xx}}, \label{fermimfp}
\eeq
 In \refdisp{Ando2,Greg2}  the authors note that in a metallic system  this parameter is expected to be greater than unity, and its least value is $\langle k_F \rangle \;  \lambda_{m}=1$ for the case of unitary (impurity) scattering. Thus we expect   that $\rho_{xx} \leq \rho_0$, i.e. $\bar{\rho}_{xx} \leq 1$ in a good metal.
 The Ioffe-Regel-Mott resistivity scale used in \refdisp{badmetal,Sriram-Edward,WXD} provides a similar measure  for quantifying  the magnitudes of the resistivity found in  strongly correlated metals.  However we should keep in mind  that   both estimates suffer from    ambiguities  in  defining a precise   threshold  value of resistivity, since factors of 2 (or of $2 \pi$) cannot be ruled out in \disp{fermimfp}.

\begin{figure}
\subfigure[ \; $\delta=0.18$. t'/t marked above]{\includegraphics[width=.475\columnwidth]{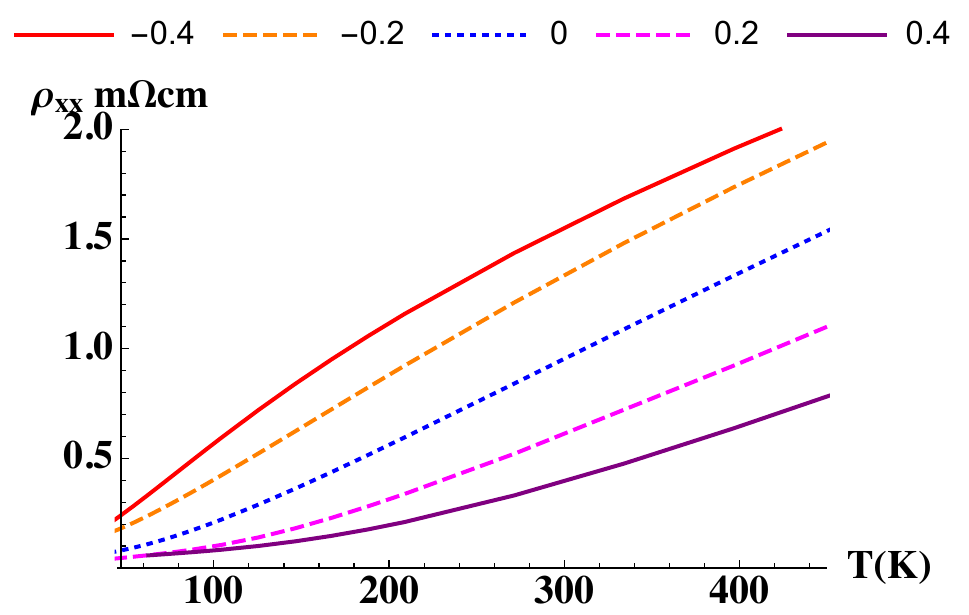}}
\subfigure[\;   $\delta=0.15$.  t'/t  marked in inset and  above]{\includegraphics[width=.475\columnwidth]{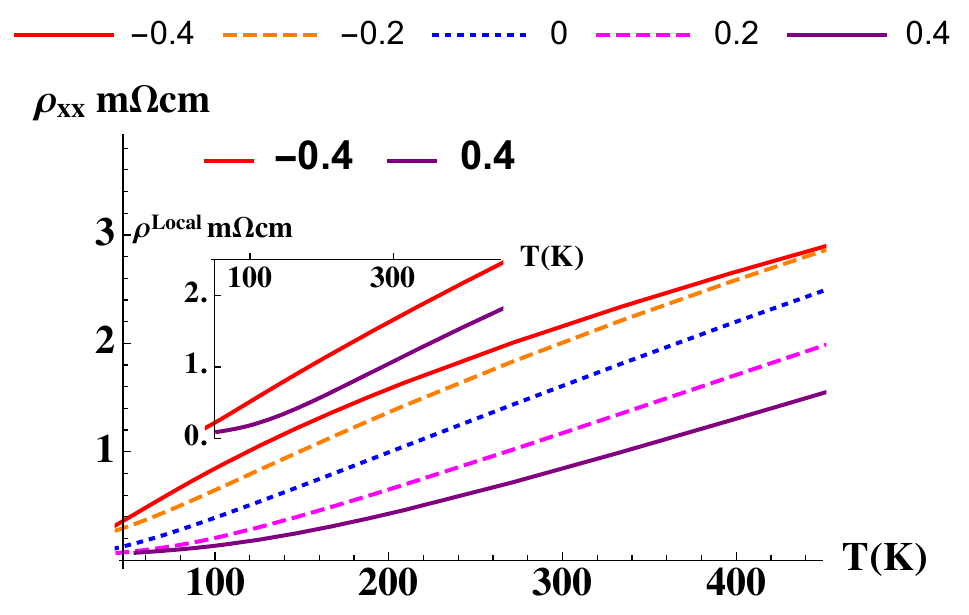}}
\subfigure[\;  $\delta=0.12$. t'/t marked above]{\includegraphics[width=.475\columnwidth]{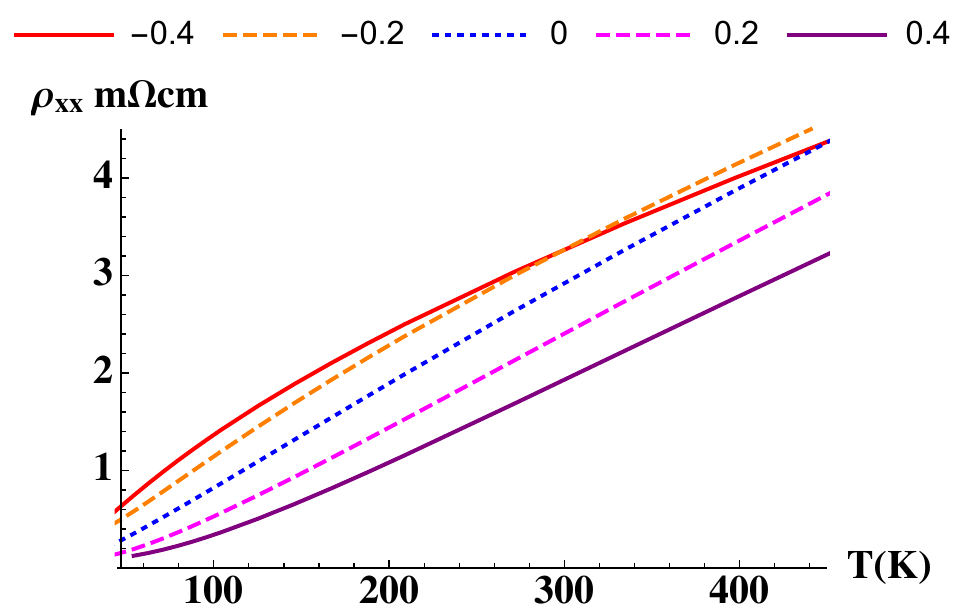}}
\caption{ \small Panels (a,b,c) show the resistivity at three densities.  We expect the very low-T region is  cutoff by superconductivity.
 Panel (b) (Inset) is  the local approximation from \disp{local}. It illustrates the  $t'$ dependence of $\rho^{Local}$ from self energies,  further enhanced by  velocity factors in the full curve.
 The  Fermi liquid $\rho \propto T^2$ regime is   shrunk (enlarged)  as $t'/t\to -0.4$ ($+0.4$). For a fixed T the curvature changes from positive to negative as $t'/t$ varies upwards in each panel, and also as n increases  across the panels -  resembling  the experimental findings of \refdisp{Ando,Sam-Martin,Takagi,NCCO-2,Greven}.  Note that the scale of the resistivity exceeds the approximately estimated unitarity value $1.71 m\Omega$cm  at modest  T for  $t'\leq0$, indicating very strong inelastic scattering. In the displayed range, the case  $t'>0$ shows a somewhat smaller resistivity, and crosses $1.71 m\Omega$cm only at the lowest hole density $\delta=0.12$.
\label{Resistances} }
\end{figure}

\figdisp{Resistances} shows the temperature dependence of the resistance at three densities, and their strong variation with $t'/t$.   $J$ is taken as $900$K here, varying J between $0$ and $1500$K makes almost no difference at these temperatures.   We see that the scale of the resistivity for $t'\leq 0$  exceeds the (approximately estimated) unitarity value $1.71 m\Omega$cm already at modest  T, indicating very strong inelastic scattering. On the other hand $t'>0$ shows a considerably smaller resistivity at most densities.

 In all curves we see that the curvature changes from positive (for $t' \geq 0$) to negative (for $t' <0$) at say 150K.  To understand the role of $t'/t$ we note that
 the resistivity in \disp{parallel}  depends  on  $t'/t$ through the velocity $v_k^x$, in addition to a dependence through  the self energies Eqs.~(\ref{eq3},\ref{eq4}).  To  gauge their relative importance it is useful to examine a local approximation  of \disp{parallel} where the two functions are averaged separately over momentum:
 \beq
\bar{\sigma}^{local}_{xx} = \langle \Upsilon(\vec{k}) \rangle_k  \times \langle ({\hbar v_k^x})^2/{a^2_0}\rangle_k. \label{local}
\eeq
The   velocity squared  average is independent of the sign of $t'$, therefore the local approximation, shown  in the inset of Panel(b),   probes only   the  dependence through Eqs.~(\ref{eq3},\ref{eq4}).  Comparing the inset  and main figure in Panel(b),
 we see that at $t'=0.4t$ both resistivity curves display a positive curvature. At $t'=-0.4 t$ we see that $\rho^{Local}$ is essentially linear in T, while $\rho_{xx}$ shows a negative curvature.
The behavior of $-\Sigma''$ in the inset of \figdisp{Fig2} qualitatively  resembles  the resistivity.  The difference is actually  related to the velocity factors, which are very different effect between $t'<0$ and $t'>0$. These   cause  the integrals to have  very different thermal variation.

  \S {\bf Hall response:}  Within the  bubble scheme,  we may also calculate the Hall conductivity\cite{voruganti,hall-extra,Tremblay,HFL} as $ \sigma_{xy}= - 2 \pi^2/{\rho_0} \times (\frac{\Phi}{\Phi_0}) \times \; \bar{\sigma}_{xy} $,   the dimensionless conductivity:
  \beq 
 \bar{\sigma}_{xy}&=& \frac{4 \pi^2 }{3 }   \int_{-\infty}^\infty d\omega \, (- {\partial f}/{\partial \omega}) \langle \rho^3_G(k,\omega) \eta(k) \rangle_k, \;\;\;\; \label{hall1}
 \eeq
 and $\eta(k)=\frac{ \hbar^2}{  a_0^2} \{ (v_k^x)^2  \frac{\partial^2 \varepsilon_k}{\partial k_y^2}- (v_k^x v_k^y) \frac{\partial^2 \varepsilon_k}{\partial k_x \partial k_y} \}$. Here    $\Phi=B a_0^2$ is  the  flux\cite{lattice}   and $\Phi_0= hc/(2 |e|)$ is the flux  quantum. In terms of these we can compute the  Hall constant $R_H$ and   Hall angle $\Theta_H$  from
 \beq
 c \, R_H & =& - \frac{4\pi^2 v_0}{|e|}  \; \bar{\sigma}_{xy} \times \bar{\rho}_{xx}^2, \label{dimensionless-Hall} \\
 \cot(\Theta_H)&=& - \frac{1}{2 \pi^2}  \frac{\bar{\sigma}_{xx}}{\bar{\sigma}_{xy}} \times \frac{\Phi_0}{\Phi}, \label{cot-Hall}
 \eeq
 with $v_0=(a_0^2 c_0)$ \cite{lattice,norms-hall}.
 
\begin{figure}[h]
\subfigure[\;\; $\tan\Theta_H$, $\delta$  marked above ]{\includegraphics[width=.5\columnwidth]{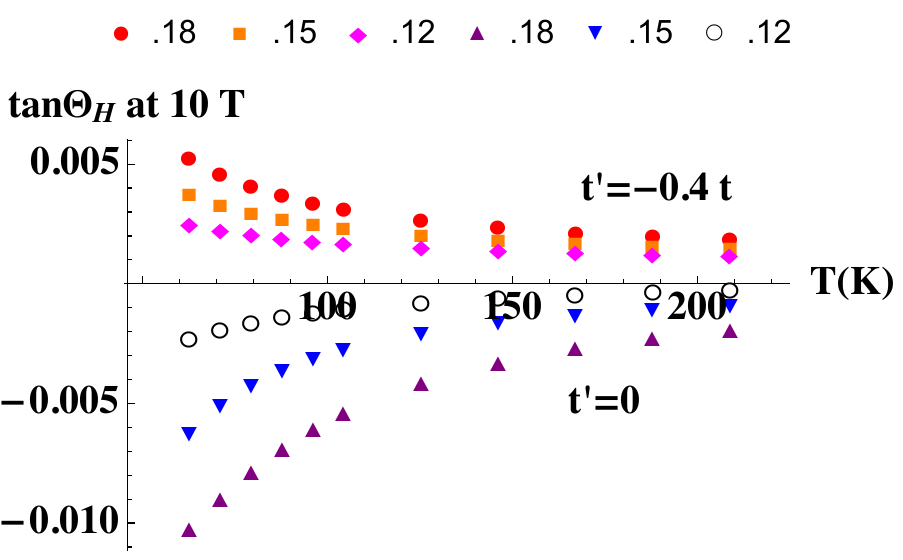}}
\subfigure[ $|\cot\Theta_H|$,  $\delta=0.15$. t'/t marked above]{\includegraphics[width=.5\columnwidth]{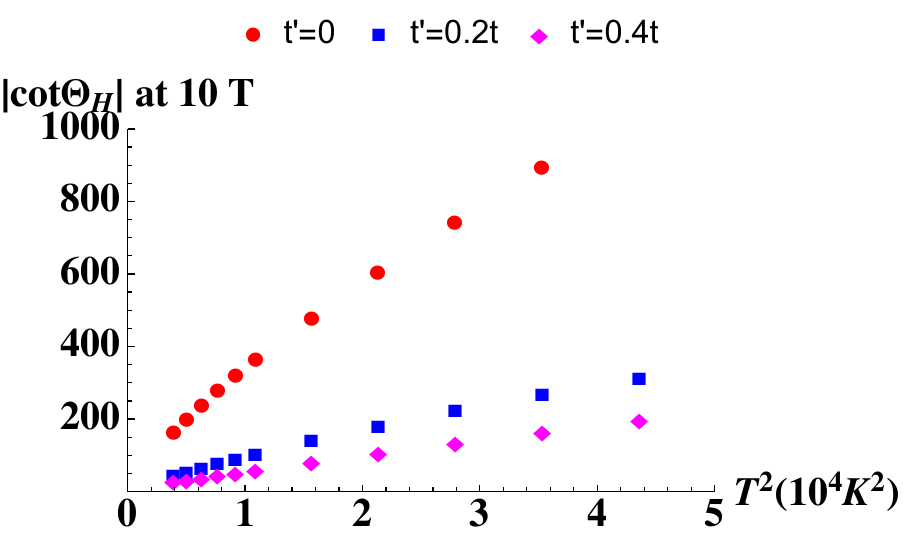}}
\subfigure[ $R_H$\ full,dotted,dashed lines are at $\delta=0.18,0.15,0.12$.  t'/t marked above]{\includegraphics[width=.5\columnwidth]{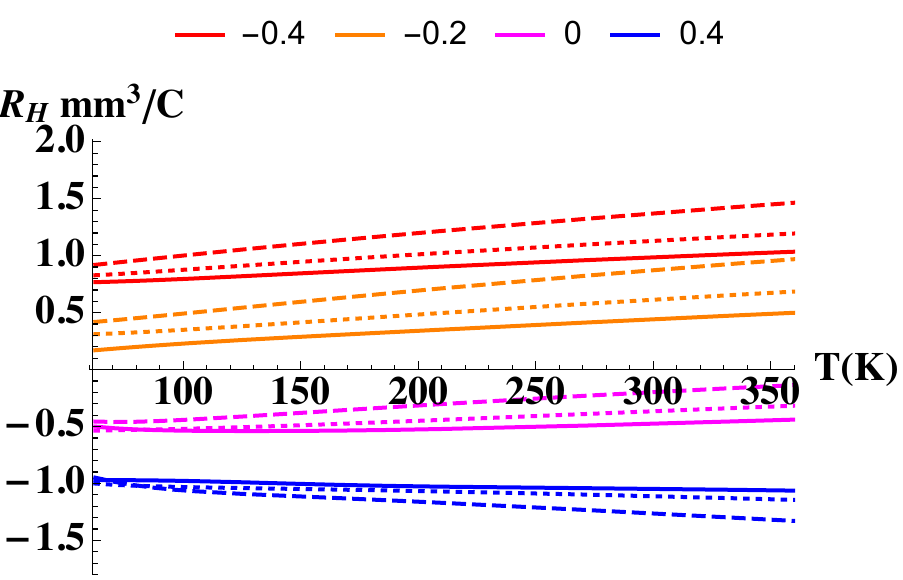}}
\caption{ \small Panel (a) $\tan\Theta_H$ for $B=10$T at three densities vs T. The   set with  $t'=-0.4t$ have a hole-like FS while the  set with $t'=0$  an electron-like FS. In both cases we see a rapid fall-off with T,  and a decreasing magnitude with $\delta$.
Panel(b) for $B=10$T at $\delta=0.15$  shows  $|\cot (\Theta_H)|$ for three values of  $t'/t$.
It is approximately   linear  with $T^2$ over the range,  in fact it is linear on both sides of a bend \cite{Takagi,Hwang,ong,ong-anderson,Ding2,Ando-Hall,Boebinger,Greven}. Panel(c) gives the T dependent $R_H$ for three densities, each with four values of $t'/t$. The sign change resembles the change seen in experiments\cite{NCCO-Hall}.
 \label{Hall} }
\end{figure}

In \figdisp{Hall} we display the computed Hall variables. In Panel (a)  $\tan\Theta_H$  is shown  for two values of $t'/t$ displaying hole-like and electron-like behavior. A decrease in hole  density reduces the magnitude in either case. 
 In Panel (b) we display the computed $\cot (\Theta_H)$ versus $T^2$ with  three values of $t'$ giving an electron-like FS. We note  that $\cot{\Theta_H}$ is  approximately linear with $T^2$ \,\cite{ong,ong-anderson,Ding2} and is strongly affected by the magnitude of $t'$. {The two distinct $\cot(\Theta_H)\propto T^2$ regimes seen in  Fig.~(\ref{Hall}-b) are also seen in many experiments, the crossover to a strange metal corresponding to the bending temperature\cite{Ding2}.} Our results for the hole-like case $t'<0$ at low $T$ are less robust due to  the small magnitude of $\sigma_{xy}$ and errors from the oscillating  sign of  $\eta$ in \disp{hall1}. In Panel (c) we show the Hall constant $R_H$ at three densities  for  representative values of $t'/t$.   Its sign is electron-like  for $t'>0$ and hole-like for $t'\leq 0$, tracking  the change in  topology of the Fermi surface in \figdisp{Fig2}.  The  magnitude  of $R_H$ is  substantially affected by changing  $t'$. This  is a strong correlation effect, and   discourages envisaging  any simple relationship between the Hall number and  hole density.
\section{Discussion}
Using the recently developed  second order equations of the ECFL theory in \refdisp{Sriram-Edward}, we have presented results for the 2-d \tJ model at low and intermediate temperatures. In keeping with our recent findings for the $d=\infty$ solution of {\em the same equations}, we note that the quasiparticle weight $Z(k_F)$ is non-zero, but  remarkably small. This fragile Fermi liquid therefore has an extremely low effective Fermi temperature,  above which it displays characteristics of a   Gutzwiller correlated strange metal, as listed in \refdisp{Sriram-Edward,WXD}, including a resistivity that is linear in T.

 By varying t', the second neighbor hopping at a fixed t and J, we found in \figdisp{Fig1} a remarkable variation of the quasiparticle weight $Z(k_F)$ that is characteristic of the 2-d square lattice, with no simple analog in $d=\infty$. We found  $t'<0$ leads to a considerable reduction in its  magnitude, while $t'>0$ leads to a larger value and thus a more robust Fermi liquid. A direct calculation of the single particle spectral width $\Gamma= -Z \Sigma''$ confirms this observation in \figdisp{Fig2}, and when studied as a function of the temperature,  shows a much larger magnitude, and hence broader  spectral lines.
 
 Our two striking results concern the spectral heights over the Brillouin zone, and  the resistivity as a function of T at various densities and t'.   The spectral height is the peak value of $\rho_{\G}(\vec{k}, \omega)$ scanned over $\omega$,  and equals the inverse of the least magnitude of  $\Im m \Sigma(\vec{k},\omega)$.  In \figdisp{Fig3} we present both the T evolution (going horizontally) and the t' evolution (going vertically) of this important object visible in ARPES. We note that $t'<0$ model with a very small $\Gamma$ also displays a rapid loss of coherence on warming. The quasiparticle peaks drop rapidly, while the valleys, representing the background spectral weight in photoemission, catch up with the peaks in magnitude. A similar variation happens for $t'=0$ but the  drop of the peak heights is more pronounced. The case of $t'>0$ has  the largest drop of peak heights, while  its effective Fermi temperature is the largest of the three cases. 
It follows that the electron doped case has a more robust Fermi liquid appearance for T lower than its Fermi scale. 
 Our study provides absolute scale values for these observable  heights, and it should be interesting to study these experimentally for comparison. Towards that objective we note that $t'>0$ maps to the electron doped High Tc superconductors, while $t'\leq 0$ maps to the hole doped cases, as we may also deduce from the shapes of the Fermi surfaces in the above figure.

 The other striking results concerns the resistivity. We are able to calculate the longitudinal resistivity $\rho_{xx}$  on a {\em doubly absolute scale}, both the magnitude of   $\rho_{xx}$ and that of T are given in physical units by using reasonable values for the basic parameters of the \tJ model and the lattice constants \figdisp{Resistances}.
We find essentially the experimentally observed scales for both axes, and there is  room for further adjustments of bare scales if needed.  The main finding is that as $\delta$ is varied towards half filling, the regime of linear resistivity increases in the hole-like cases ($t'\leq 0$) and the quadratic dependence regime  shrinks to very low T scales, falling   below the known superconducting transition temperatures. The other important finding is that the concavity (convexity)  of resistance versus T, usually taken to denote a (non) Fermi liquid behavior, requires an enlarged viewpoint; we find that the sign of t'  flips the two cases. As an example, the case $t' \leq 0$ has a pronounced convex regime at low T. This could be naively ascribed to a non-Fermi liquid behavior, but in reality is a crossover range to the strange metal regime.

We also present results in \figdisp{Hall} for the Hall constant and the Hall angle. These are calculated  using simple versions of the Kubo formula, found by neglecting the vertex corrections, in the same spirit as the longitudinal resistivity.
It must be kept in mind that the vertex corrections are likely to be more significant for the Hall response, since there are two vertices involved- the second one from the magnetic field derivative of the Greens function. Additionally there are no exact results in literature on correlated matter  for the Hall constant to benchmark the ECFL results. For these reasons
 one might place  lesser confidence in the quantitative aspect of the Hall results as compared to the longitudinal transport functions.
We find  that the Hall angle  changes sign with t'. The $\cot(\Theta_H)$ is found to be roughly linear with $T^2$, in agreement with the experimental situation. Interestingly the magnitude of the computed results is also roughly in accord with the experiments. While these results are   encouraging, we believe that  further work is needed to unravel what  we appears to be a knee in the $\cot(\Theta_H)$ versus $T^2$ curve, and also  to better estimate the density dependence of the Hall constant.
 
\section { Conclusions:}  In this work, we used a scheme from the ECFL theory where  the second order $\lambda$ expansion terms are supplemented  with   a high energy  cutoff. This scheme has been benchmarked in $d= \infty$ against DMFT \cite{Sriram-Edward,WXD} for computing transport and other low energy excitations,  giving   good agreement with exact numerical results.  As detailed in \refdisp{Sriram-Edward} the magnitude of the quasiparticle weight $Z$ is somewhat lower in this scheme as compared to  the exact DMFT values for hole density $\delta \lessim 0.8$ resulting in a larger magnitude of the resistivity as well.   In this work the same formalism has been applied to the 2-d \tJ model.  It is  possible that the  close agreement found in the $d=\infty$ case might not survive to the same extent for $d=2$.  Hence    we might  expect to find further  refinements of the absolute  values of the quasiparticle weight and resistivity  to emerge from  further studies of the $\lambda$ expansion. However it seems  likely that the crucial  variation of resistivity and Hall constant with the magniude and sign of $t'$ found here will persist in more exact future results.  Hence it seems that  we can  draw some useful conclusions already regarding the difference between hole and electron doping.

 We have shown a  range of results   for the 2-d  \tJ model, obtained by varying different  parameters within  our scheme. It is interesting  that the magnitudes of  various transport variables, presented here in physical units\cite{errors}, are roughly  on the scale of reported measurements\cite{Takagi,Hwang,Ando,NCCO-2,Boebinger,Sam-Martin,Ando-Hall,Greven}.   Although it is not our primary aim here to produce exact fits, we note that  the agreement  can be improved in many cases with suitable changes of the bare  (band) parameters.

In the  range of parameters considered here,  a metallic state has been posited, and therefore the role of the exchange J is limited; we find very little variation of the transport quantities with  a change  in  $J$. The transport parameter variation with density seems very similar to that found in $d=\infty$ in \refdisp{badmetal,Sriram-Edward,WXD} where  a large variety of  Gutzwiller correlated metallic states were shown to arise\cite{WXD}, with their origin in the $U=\infty$ or  Gutzwiller correlation rather than with J. The added feature in $d=2$ is the important role played by t', as stressed here.
 We expect magnetic,  superconducting and possibly other broken symmetry states  at the lowest T and $\delta$ to arise,  largely due  to the effect of J.  Further work is necessary to find reliable  calculational schemes  for these broken symmetry states.

  A few  broad    conclusions suggest themselves.   The  parameter $t'/t$ plays a key role in  determining the low-energy scales. In \figdisp{Fig1} we see that the quasiparticle weight $Z$ has a large variation with $t'$.
 The  origin of this sensitivity lies in  the  self energies in Eqs.~(\ref{eq3},\ref{eq4}), where combinations of the band energies $\varepsilon_k$  play the role of an effective interaction. Varying $t'/t$ therefore  changes the  self-energies   strongly, in contrast   to the  usual  weak   change   via the  band parameters in  \disp{eq2}.

Our main findings are  as follows.   ({\bf I})   The spectral functions are highly sensitive to thermal variation; in \figdisp{Fig3} we observe  a five to fifteen fold  drop in intensity with a  variation of  $k_BT$ about 1/100$^{th}$    the  bandwidth $\sim 3.6$ eV. This is in severe conflict with   expectations from   conventional theories of  metals.       ({\bf II}) We note from \figdisp{Resistances}  that with  $t'\leq 0$, a Fermi liquid (FL) resistivity $\rho\propto T^2$ is seen  only at   very  low T. The very low T (FL) regime    is followed  by a ``strange metal'' regime, also at low $T$, where we find  a  $\rho\, vs \,  T$ curve, with zero or negative curvature.    This  regime parallels   the Gutzwiller-correlated strange metal regime   reported  in $d=\infty$  \cite{WXD},   the negative  curvature making it even  stranger. ({\bf III}) For   the electron-doped case $t'>0$, \figdisp{Resistances} shows  that the curvature is positive and the Fermi liquid regime  extends to  higher temperatures. 

It is significant  that  the  ECFL theory captures the diametrically opposite resistivity behaviors of  hole doped\cite{Takagi,Sam-Martin,Ando} and electron doped materials\cite{NCCO-2,Greven} within the same scheme, only differing in the sign of $t'/t$.  The resistivity curvature mapping of \refdisp{Ando} can also  be viewed in terms of a variation of this ratio and the temperature, as in \figdisp{Resistances}. In conclusion   this work
provides a sharp picture of the difference made by the second neighbor hopping $t'$ in the presence of Gutzwiller correlations. It also  yields quantitative results for  several famously  hard to compute variables in  correlated matter, that in  rough  agreement with a variety of experiments.

\section{Acknowledgements}
We thank Edward Perepelitsky and Sergey Syzranov for helpful comments on the manuscript. 
The work at UCSC was supported by the U.S. Department of Energy (BES) under Award \# DE-FG02-06ER46319. Computations reported here used the XSEDE Environment\cite{xsede} (TG-DMR160144) supported by National Science Foundation grant number ACI-1053575.


\begin{thebibliography}{99}

\bibitem{PWA} P. W. Anderson, Science {\bf 235}, 1196 (1987).

\bibitem{tJ-review1} M. Ogata and H Fukuyama, Rep. Prog. Physics, {\bf 71}, 036501 (2008); https://doi.org/10.1088/0034-4885/71/3/036501

\bibitem{ECFL}  B. S. Shastry, arXiv:1102.2858 (2011), Phys. Rev. Letts. {\bf 107},  056403 (2011).

\bibitem{Pathintegrals}  B. S. Shastry, arXiv:1312.1892 (2013),  Ann. Phys. {\bf 343}, 164-199 (2014).  DOI:http://dx.doi.org/10.1016/j.aop.2014.02.005.  (Erratum) Ann. Phys. Vol.  373, 717-718 (2016).  \\ DOI:http://dx.doi.org/10.1016/j.aop.2016.08.015.

\bibitem{Sriram-Edward}B. S. Shastry and E. Perepelitsky, arXiv:1605.08213. Phys. Rev. {\bf B 94}, 045138 (2016). DOI: http://link.aps.org/doi/10.1103/PhysRevB.94.045138; R. \v{Z}itko, D. Hansen, E. Perepelitsky, J. Mravlje, A. Georges and B. S. Shastry, arXiv:1309.5284 (2013), Phys. Rev. {\bf B 88}, 235132 (2013). DOI:http://dx.doi.org/10.1103/PhysRevB.88.235132,  B. S. Shastry, E. Perepelitsky and A. C. Hewson, arXiv:1307.3492 [cond-mat.str-el], Phys. Rev. {\bf B 88}, 205108 (2013). DOI:http://dx.doi.org/10.1103/PhysRevB.88.205108.

\bibitem{Edward-Sriram} E. Perepelitsky and B. S. Shastry, Ann. Phys. {\bf 357}, 1 (2015). DOI:
http://dx.doi.org/10.1016/j.aop.2015.03.010

\bibitem{WXD}  W.  Ding, R. \v{Z}itko, P. Mai, E. Perepelitsky and  B. S. Shastry, arXiv:1703.02206v2, Wenxin Ding, Rok \v{Z}itko, and B. Sriram Shastry, arXiv:1705.01914

\bibitem{notation-1}  We denote $k\equiv(\vec{k}, i \omega_n)$, $\omega_n=  (2n+1) \pi k_B T$  the Matsubara frequencies, $N_s$ the number of sites and  $\sum_k \equiv \frac{k_B T}{N_s} \sum_{k_x,k_y, \omega_n} $. $J_k$ is the Fourier transform of the  exchange.

\bibitem{badmetal} X.Y. Deng, J, Mravlje, R. \ifmmode \check{Z}\else \v{Z}\fi{}itko, M. Ferrero, G. Kotliar and A. Georges, Phys. Rev. Lett. {\bf 110}, 086401 (2013).

\bibitem{HFL} W. Xu, K. Haule, and G. Kotliar, Phys. Rev. Lett. {\bf 111}, 036401 (2013).

\bibitem{Monster}  B. S. Shastry, arXiv:1207.6826 (2012); Phys. Rev. {\bf B 87}, 125124 (2013).



\bibitem{Hansen}   D. Hansen and B. S. Shastry, Phys. Rev. {\bf 87} 245101 (2013). 


\bibitem{u0}  Observe that in these equations, an arbitrary  shift  of the  band $\varepsilon_k \to \varepsilon_k +c$   can be absorbed into $u_0$. Thus the {\em shift invariance}  is manifest to second order in $\lambda$.


\bibitem{exptl-tprime} In high Tc systems  \refdisp{tJ-review1} estimate  $t'\lessim -.27$  for BSCCO, while for LSCO  $t'\sim -0.16 t$. NCCO is modeled with $t'>0$ after invoking a particle hole transformation. In this case  we must  flip  the sign of the calculated $R_H$ and $\Theta_H$  to compare with data. 

\bibitem{Bansil}  R. S. Markiewicz, S. Sahrakorpi, M. Lindroos, Hsin Lin, and A. Bansil, Phys. Rev. {\bf B 72}, 054519 (2005).    An  extended set of hopping parameters is recommended here.

 


\bibitem{bubble} B. S. Shastry and B. Shraiman, Phys. Rev. Letts. {\bf 65}, 1068 (1990). 


\bibitem{lattice}
The numerics assume a bct unit cell $(a,a,c)$ with $a=3.79\, \AA^0$ and $c=13.29 \, \AA^0$.  In the expression for $\rho_0$, $c_0$ corresponds to the  interlayer separation $c_0=c/2$.
 In \disp{dimensionless-Hall} and  \disp{cot-Hall}  we use $v_0/|e|= .596\times10^{-3} \,cm^3/C$   and  $\Phi_0/\Phi= 1440$ with $B=10T$. 

 

\bibitem{Ando}Y. Ando, S. Komiya, K. Segawa, S. Ono, and Y. Kurita, Phys. Rev. Letts. {\bf 93}, 267001 (2004). 

\bibitem{Sam-Martin} S. Martin, A. T. Fiory, R. M. Fleming, L. F. Schneemeyer
and J. V. Waszczak, Phys. Rev. {\bf B 60}, 2194 (1988).


\bibitem{Takagi} H. Takagi, T. Ido, S. Ishibashi, M. Uota,  S. Uchida, and Y. Tokura, Phys. Rev. {\bf B 40} 2254 (1989).

\bibitem{NCCO-2}Y. Onose, Y. Taguchi, K. Ishizaka, and Y. Tokura, Phys.
Rev. {\bf B 69}, 024504 (2004).

\bibitem{Greven} Y. Li, W. Tabis, G. Yu, N. Bari\v{s}i\'c and M. Greven, Phys. Rev. Letts. {\bf 117}, 197001 (2016).

\bibitem{voruganti}
P. Voruganti, A. Golubentsev and S. John, Phys. Rev. {\bf B 45}, 13945 (1992);  H. Fukuyama, H. Ebisawa, and Y. Wada: Prog. Theor. Phys. {\bf 42} 494 (1969); H. Kohno and K. Yamada, Prog. Theor. Phys. {\bf 80} 623 (1988);


\bibitem{hall-extra} For this we additionally assume that the magnetic field vertex also assumes  its  bare value. This assumption requires further  validation in 2-dimensions within the \tJ model,  hence the results for the Hall conductivity are less reliable than the longitudinal conductivity.


\bibitem{Tremblay} L-F Arsenault and A.M. S. Tremblay
Phys. Rev. {\bf B 88}, 205109 (2013)




\bibitem{norms-hall} These definitions lead to intuitive  results in a simple case.  For  2-d electrons with  $\varepsilon_k = \hbar^2 k^2/(2 m)$, and  a Lorentzian  $\rho_G(k,\omega)$ of width $\Gamma$,   we recover the Drude result $\sigma_{xx} = n q_e^2 \tau/m$ and $\bar{\sigma}_{xy}=  \frac{n}{2 \pi^2} (\frac{\hbar^2}{2 m a_0^2 \Gamma})^2$,  where $n$ is the number of electrons per site, and
 $\tau= \hbar/(2 \Gamma)$. Thus  $|e| R_H c/v_0= -1/n$ in  \disp{dimensionless-Hall}, and $\cot(\Theta_H)= -1/(\omega_c \tau)$ in  \disp{cot-Hall}   where $\omega_c\equiv |e| B/(m c)$. 

\bibitem{ong} T. R. Chien, Z. Z. Wang and N. P. Ong, Phys. Rev. Letts. {\bf 67}, 2088 (1991). 

\bibitem{ong-anderson}N. P. Ong and P. W. Anderson, Phys. Rev. Letts. {\bf 78}, 977 (1997).

\bibitem{Ding2} Wenxin Ding, Rok \v{Z}itko, and B. Sriram Shastry, arXiv:1705.01914. See Fig.3 and related discussion.
 
\bibitem{errors} From \refdisp{Sriram-Edward,WXD} we may infer that the $Z$ in the present calculation is a bit too low for $t'=0$ and $ .12\leq  \delta \leq .15$. This  is expected to result in overestimating $\rho_{xx}$ by a factor $\sim 3$ at $T=450$K.


\bibitem{Hwang} H. Y. Hwang, B. Batlogg, H. Takagi, H. L. Kao, J. Kwo, R. J. Cava, J.J. Krajewski and W. F. Peck, Jr., Phys. Rev. Letts. {\bf 72} 2636 (1994).

\bibitem{Ando-Hall} Y. Ando, Y. Kurita, S. Komiya, S. Ono and K. Segawa, Phys. Rev. Letts. {\bf 92}, 197001 (2004).

\bibitem{Boebinger}F. F. Balakirev, J. B. Betts, A. Migliori, I. Tsukada, Y. Ando, and G. S. Boebinger, Phys. Rev. Letts. {\bf 102}, 017004 (2009).
\bibitem{NCCO-Hall} J. Takeda,T. Nishikawa, M. Sato, Physica {\bf C 231}, 293 (1994).  See esp. Fig.~(4).  


 \bibitem{xsede} J. Town et al., ``XSEDE: Accelerating Scientific Discovery", Computing in Science \& Engineering, Vol.16, No. 5, pp. 62-74, Sept.-Oct. 2014, doi:10.1109/MCSE.2014.80
      
 \bibitem{Ando2} Y. Ando see Eq.~(6) in {\em High T$_c$ Superconductors and Related Transition Metal Oxides}, Editors: A. Bussman-Holder, H. Keller (Springer Verlag Heidelberg, 2007). The origin of this formula is simple to understand, 3-d conductivity is written in terms of the two dimensional density as $\sigma= \frac{n_{2d} e^2 \tau}{c_0 m}$, and writing $n_{2d}= k_F^2/(2 \pi)$ and $\lambda_{m}= \tau \hbar k_F/m$ we obtain $\sigma= e^2/(h c_0) k_F \lambda_{m}$.
 
 \bibitem{Greg2}   Y. Ando, G.S. Boebinger, A. Passner, T. Kimura and K. Kishio, Phys. Rev. Letts. {\bf 74} 3253 (1995). 
 

\end{thebibliography}
\end{document}